\documentclass[journal]{IEEEtran}
\usepackage{cite}
\usepackage{amsmath,amssymb,amsfonts}
\usepackage{mathtools}
\usepackage{algorithmic}
\usepackage{graphicx,color}
\usepackage{textcomp}
\usepackage{bm}
\usepackage{threeparttable}
\usepackage{colortbl}
\usepackage{booktabs}
\usepackage{subcaption}
\usepackage{multirow}
\usepackage[ruled,vlined]{algorithm2e}
\usepackage{nicefrac}
\usepackage{balance}
\usepackage{float}

\begin{document}

\title{6G WavesFM: \\A Foundation Model for Sensing, \\Communication, and Localization}

\author{\IEEEauthorblockN{Ahmed Aboulfotouh, Elsayed Mohammed, and Hatem Abou-Zeid}\\
\IEEEauthorblockA{{Department of Electrical and Software Engineering}, {University of Calgary}, Canada}\\
\texttt{ahmed.aboulfotouh@ucalgary.ca}\\
\thanks{This work has been submitted to the IEEE for possible publication. Copyright may be transferred without notice, after which this version may no longer be accessible.}}

\maketitle



\begin{abstract}

This paper introduces WavesFM, a novel Wireless Foundation Model (WFM) framework, capable of supporting a wide array of communication, sensing, and localization tasks. Our proposed architecture combines a shared Vision Transformer (ViT) backbone with task‑specific multi‑layer perceptron (MLP) heads and incorporates Low‑Rank Adaptation (LoRA) for parameter‑efficient fine‑tuning. This design promotes full parameter sharing across tasks, significantly reducing the computational and memory footprint without sacrificing performance.
The model processes both image‑like wireless modalities, such as spectrograms and channel state information (CSI), and in‑phase and quadrature (IQ) signals arranged as orthogonal frequency‑division multiplexing (OFDM) resource grids.
We demonstrate the strong generalization capabilities of WavesFM through extensive experiments on four downstream tasks: Fifth Generation New Radio (5G NR) positioning; multiple‑input multiple‑output OFDM (MIMO‑OFDM) channel estimation; human activity sensing; and radio‑frequency (RF) signal classification. Compared to supervised baselines trained individually, our approach achieves superior performance while sharing $80\%$ of its parameters across tasks.
Furthermore, we show that pretraining on domain‑relevant data not only boosts performance but also accelerates convergence, reducing training time by up to 5×. These results demonstrate that our unified WFM can support diverse tasks and deliver significant gains in both performance and efficiency, highlighting the transformative potential of foundation models to drive AI‑native paradigms in future sixth‑generation (6G) networks.

\end{abstract}

\begin{IEEEkeywords}
6G, Foundation Models, Sensing, Localization, MIMO Channel Estimation
\end{IEEEkeywords}

\maketitle

\section{Introduction} 

\begin{figure*}[h!]
    \centering
    \includegraphics[width=0.85\linewidth, keepaspectratio]{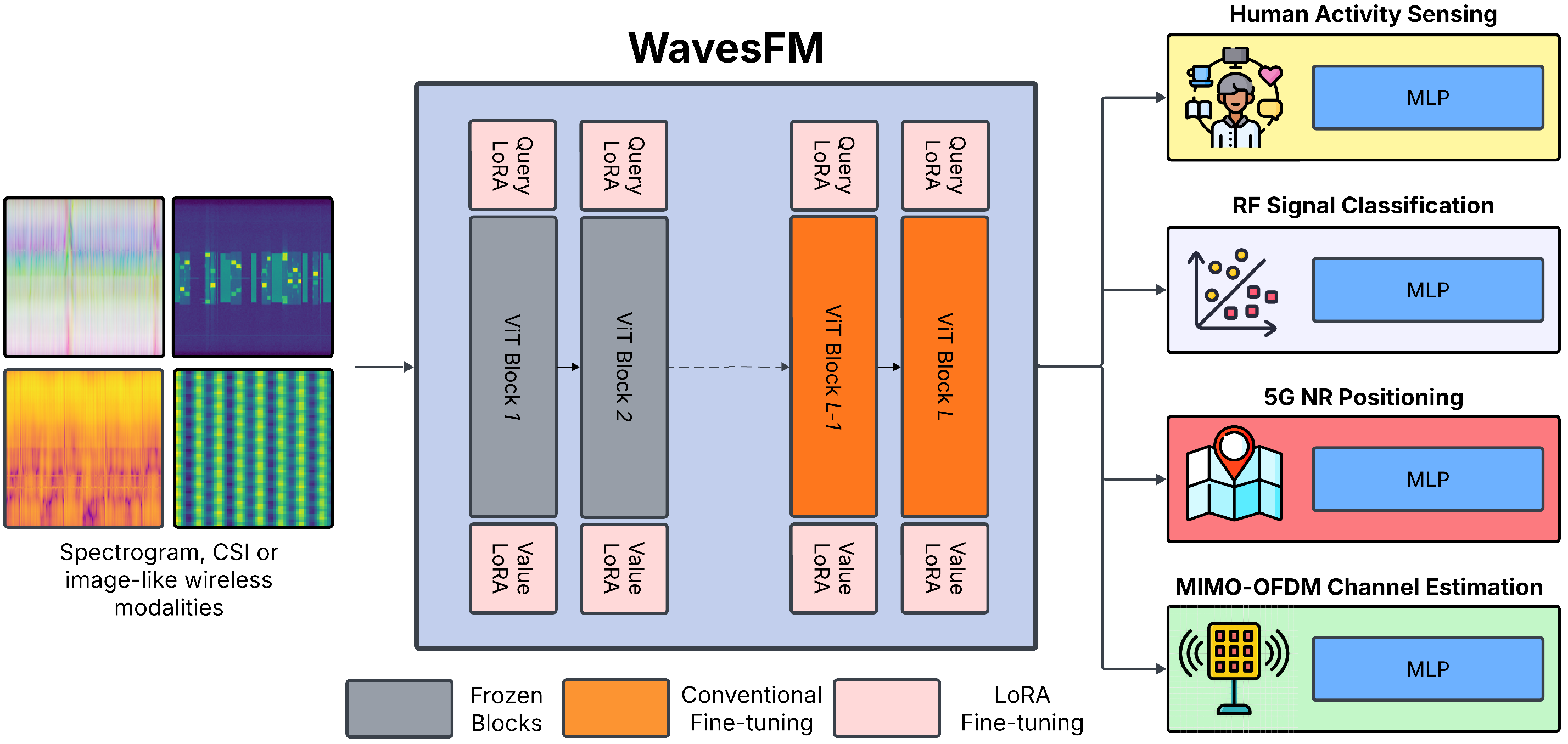}
    \caption{Overview of WavesFM, its multi-task capabilities and fine-tuning techniques.}
    \label{fig:rfm_ojcoms}
\end{figure*}

Recent developments in artificial intelligence (AI), particularly deep learning (DL), are prompting further research into how these advances can impact the evolution of 6G technologies. It is envisioned that AI will permeate every aspect of wireless networks, from design and deployment to operation, supporting a variety of network functions, enabling new services, and adapting dynamically to the environment. Enhanced adaptability and programmability through AI will also facilitate autonomous network management, allowing networks to operate with limited human intervention \cite{ITU2023M2160}.

DL has shown great potential when applied to individual wireless tasks, including automatic modulation classification \cite{amc_1, amc_2}, channel estimation \cite{chan_estim_1}, constellation and waveform design \cite{waveform_1}, among others, most of which rely on supervised learning (SL). Early success of DL came from SL, which involves collecting a labeled dataset for a specific task, that is used to train and evaluate the model. However, SL has several drawbacks. Not every task has a labeled dataset, and labeling is costly and time-consuming. Moreover, the resulting models are dedicated for one task. In practical wireless systems with resource and latency constraints, deploying a single model for each task constitutes a huge overhead. Furthermore, SL models are highly specialized, and there are concerns about their ability to generalize effectively in real-world scenarios. Wireless signals are subject to time-varying impairments, and the communication environment is constantly changing, which can degrade a DL model’s performance if it fails to adapt. Even small changes to the input distribution or a slight modification in the system might require full retraining.

Wireless Foundation Models (WFMs) are emerging as a solution to address these challenges. Generally, a Foundation Model (FM) is an AI model with a general understanding of a specific domain. This makes it possible to adapt, or fine-tune the model to various downstream tasks within that domain. There are two phases for creating FMs: pre-training and fine-tuning. 
In the pre-training phase, the model is trained on large and diverse unlabeled data using self-supervised learning (SSL). 
In the SSL paradigm, the model learns from the structure of the data which provides the supervision \cite{ssl_survey, ssl_wireless, ericsson_self-supervised_2022}. Not needing labels is significant, because the amount of unlabeled data available in any domain is orders magnitudes larger than labeled data. Leveraging this vast amount of information enables the FM to learn broad, generalizable and adaptable representations of its target modality.
In the finetuning phase, the model is adapted to a specific task using a few task-specific labeled examples.
A WFM has a more general knowledge and internal representation of wireless modalities, which enables it to perform multiple tasks, thereby reducing both computational footprint and processing overhead. Its adaptable representations also enhance its robustness, making it well-suited for handling diverse and dynamic wireless environments, and dealing with domain shifts effectively.

\subsection{Contributions}

The primary contribution of this paper is WavesFM illustrated in Figure \ref{fig:rfm_ojcoms}. We present the first WFM capable of supporting a variety of sensing, communication and localization tasks. 
These tasks include both classification tasks, human activity sensing and RF signal identification, and regression tasks, 5G NR positioning and MIMO-OFDM channel estimation. The system combines a shared WFM backbone with task-specific multi-layer perceptron (MLP) heads for each task. The framework enables fine-tuning parts of the WFM backbone when necessary, though we encourage maximum parameter sharing across tasks. To achieve this, we employ low-rank adaptation (LoRA), a parameter-efficient fine-tuning technique that enables task-specific adaptation while keeping the WFM entirely frozen. This work builds upon and extends our prior work in \cite{aboulfotouh2024building6gradiofoundation} by pretraining with a breadth of real-world datasets that enable diverse tasks, detailed methodology, efficient LoRA fine-tuning, and comprehensive experiments and analysis.
In more detail, our contributions are:
\begin{itemize}
    \item We propose WavesFM, a Vision Transformer (ViT)–based model that learns rich wireless signal representations via self‑supervised learning. We adopt a Masked Wireless Modeling (MWM) approach for pre-training on wireless data, where portions of the input are masked and the model is trained to reconstruct the original input from a partial observation. This process enables the model to learn the underlying data distribution without relying on labels.
    \item We demonstrate that WavesFM effectively learns features that generalize across sensing, communication and localization tasks. Compared to a supervised learning (SL) baseline trained individually for each task, WavesFM outperforms SL in RF signal classification, 5G NR positioning, and MIMO-OFDM channel estimation. Notably, it achieves half the positioning error of the SL model while sharing $80\%$ of its parameters across the four tasks.
    \item We investigate the impact of pre-training data size and its relevance to downstream tasks by leveraging multiple data sources, including spectrograms, 5G CSI, and WiFi CSI. Our findings show that pre-training on data closely aligned with the downstream task significantly accelerates convergence during fine-tuning. For example, in the case of human activity sensing, it results in a five-fold reduction in convergence time compared to SL. 
    Although increasing the size of pre-training data is generally beneficial, the choice of data is both critical and nuanced: misalignment between pre-training and downstream tasks can negatively affect performance, a phenomenon known as negative knowledge transfer.
    \item Conventional fine-tuning adapts entire blocks of the model for each task, which limits parameter sharing across tasks. To enable full parameter sharing, we adopt LoRA, a technique that preserves the shared model by keeping its primary parameters frozen while introducing a small set of task-specific parameters. This allows for more fine-grained and efficient adaptation. In our 
    experiments, LoRA achieves superior performance using approximately $1.5$ million task-specific parameters, significantly fewer than the $7$ million required by conventional fine-tuning, to reach comparable accuracy.
\end{itemize}
Our results confirm that WavesFM can efficiently generalize across sensing, communication and localization tasks, offering substantial improvements in both performance and resource utilization. This approach points to a promising direction for the design of intelligent, robust and adaptable 6G networks.

\subsection{Paper Outline and Notation}

The remainder of the paper is structured as follows: Section \ref{sec:related_work} describes the related work. Section \ref{sec:datasets} presents the datasets utilized for pre-training, and the four downsteam tasks: human activity sensing, 5G NR positioning, RF signal classification, and MIMO-OFDM channel estimation.
Section \ref{sec:methods} describes the proposed WavesFM, its ViT architecture and the pre-training and fine-tuning algorithms.
Section \ref{sec:results} presents numerical experiments conducted to evaluate WavesFM. Finally, section \ref{sec:conclusion} concludes the paper.

Boldface lowercase letters, such as $\bm{x}$, denote column vectors, while boldface uppercase letters, like $\mathbf{X}$, denote matrices. Subscripts indicate indexing; for instance, $\mathbf{X}_i$ refers to the $i$-th column of matrix $\mathbf{X}$. The Euclidean norm of a vector $\bm{x}$ is expressed as $\|\bm{x}\|$.

\begin{figure*}[h!]
    \centering
    \begin{subfigure}[t]{0.33\linewidth}
        \centering
        \includegraphics[width=\linewidth,keepaspectratio]{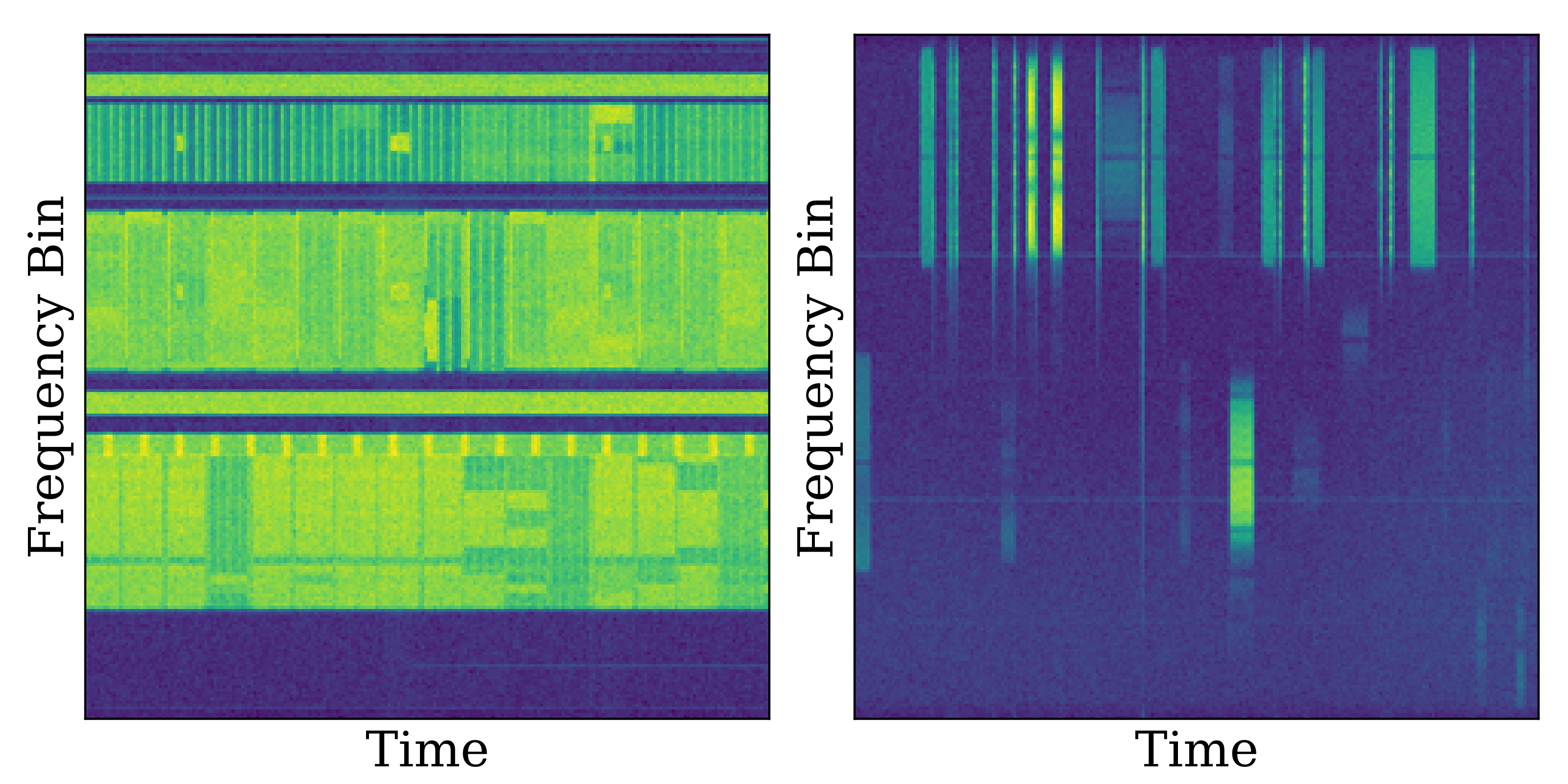}
        \caption{RF Spectrograms}
        \label{fig:pretrain_rfs}
    \end{subfigure}%
    \begin{subfigure}[t]{0.33\linewidth}
        \centering
        \includegraphics[width=\linewidth,keepaspectratio]{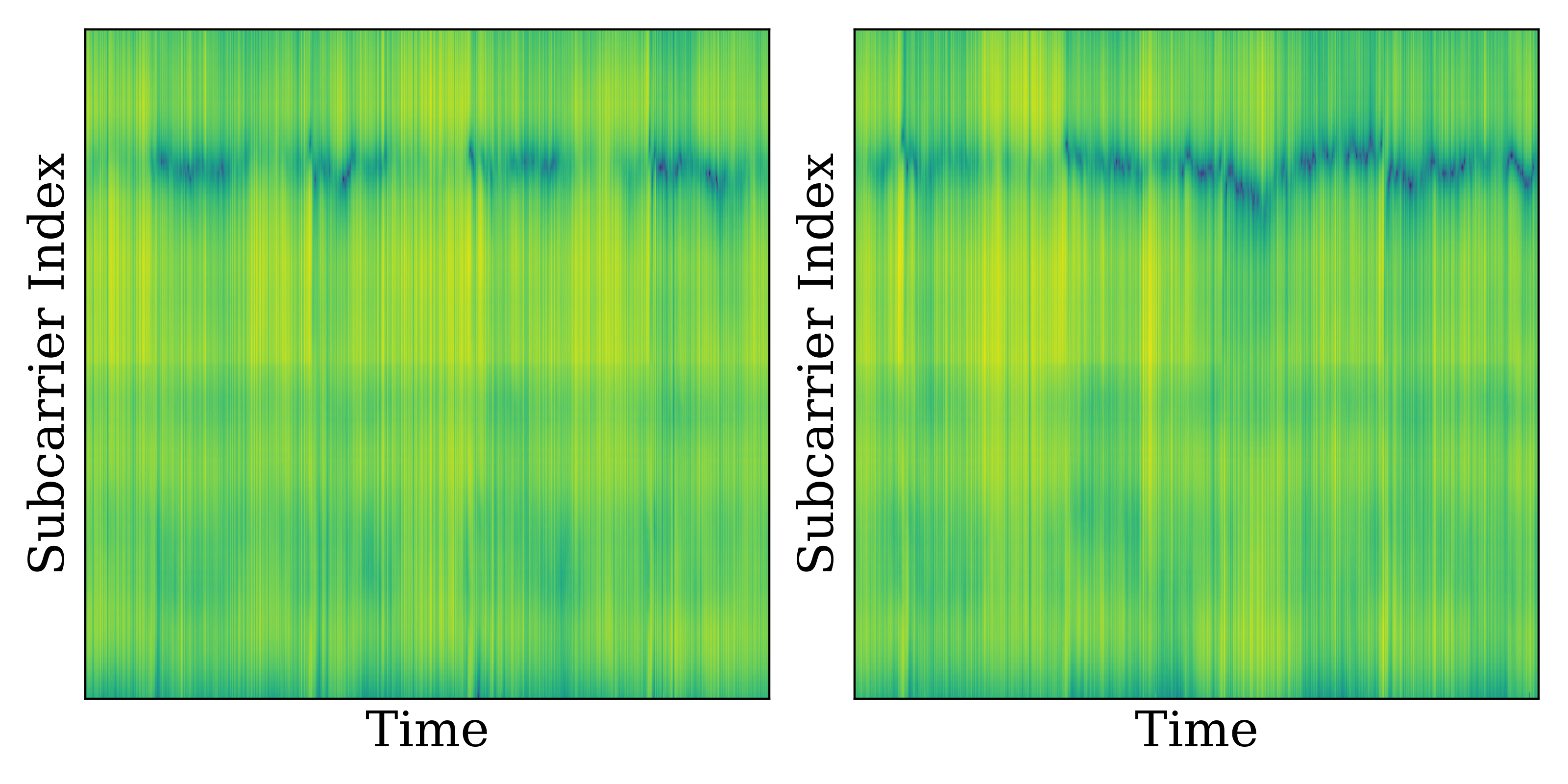}
        \caption{WiFi CSI}
        \label{fig:pretrain_wifi_csi}
    \end{subfigure}%
    \begin{subfigure}[t]{0.33\linewidth}
        \centering
        \includegraphics[width=\linewidth,keepaspectratio]{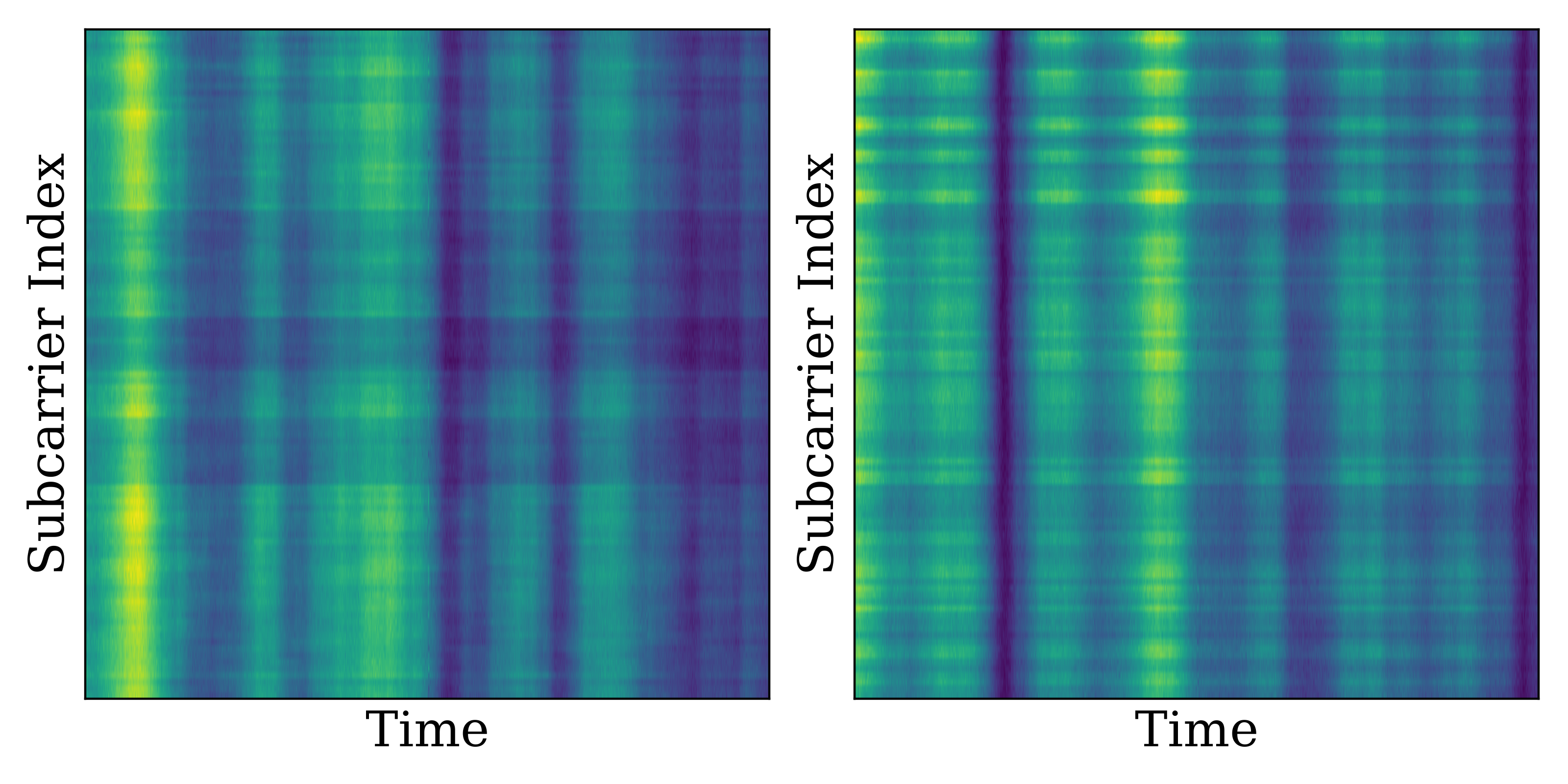}
        \caption{5G CSI}
        \label{fig:pretrain_5g_csi}
    \end{subfigure}
    \captionsetup{justification=centering}
    \caption{Samples from the pre-training datasets.}
    \label{fig:pretraining-samples}
\end{figure*}
\begin{figure*}[h!]
    \centering
    \begin{subfigure}[t]{0.2\linewidth}
        \centering
        \includegraphics[width=0.8\linewidth,keepaspectratio]{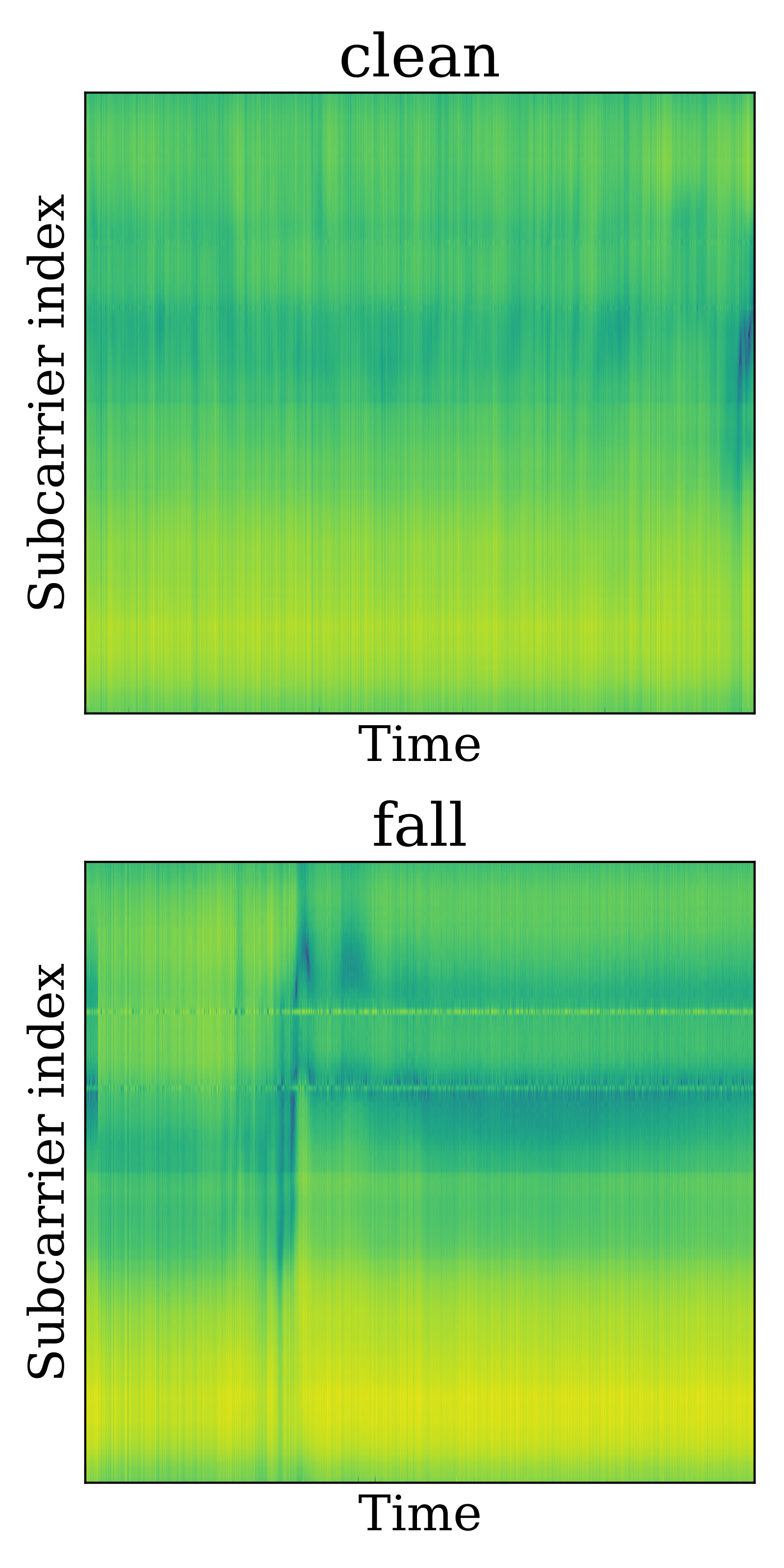}
        \caption{Human Activity Sensing}
        \label{fig:human_sensing}
    \end{subfigure}%
    \begin{subfigure}[t]{0.2\linewidth}
        \centering
        \includegraphics[width=0.8\linewidth,keepaspectratio]{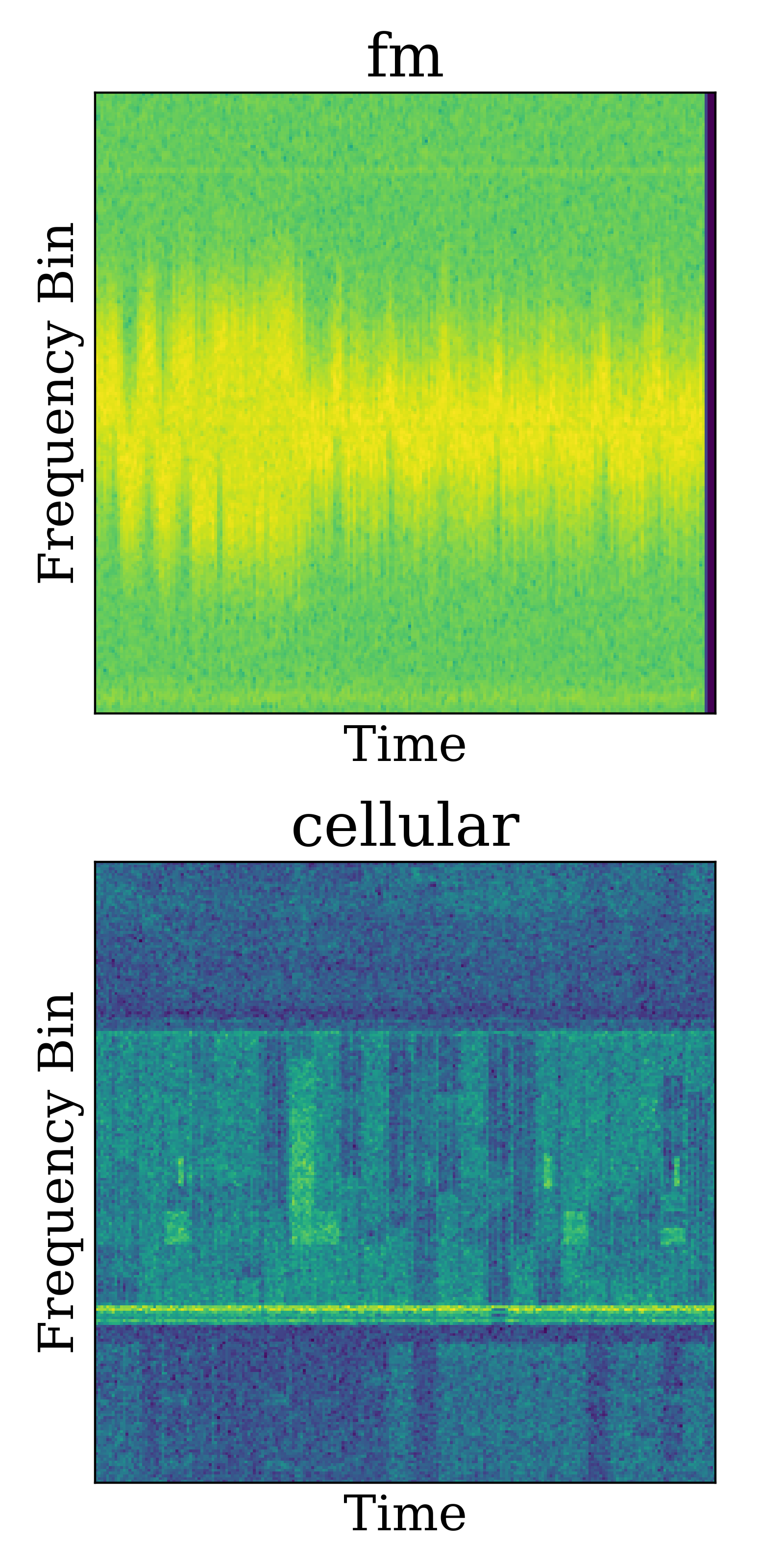}
        \caption{RF Signal Classification}
        \label{fig:rf_classification}
    \end{subfigure}%
    \begin{subfigure}[t]{0.2\linewidth}
        \centering
        \includegraphics[width=0.8\linewidth,keepaspectratio]{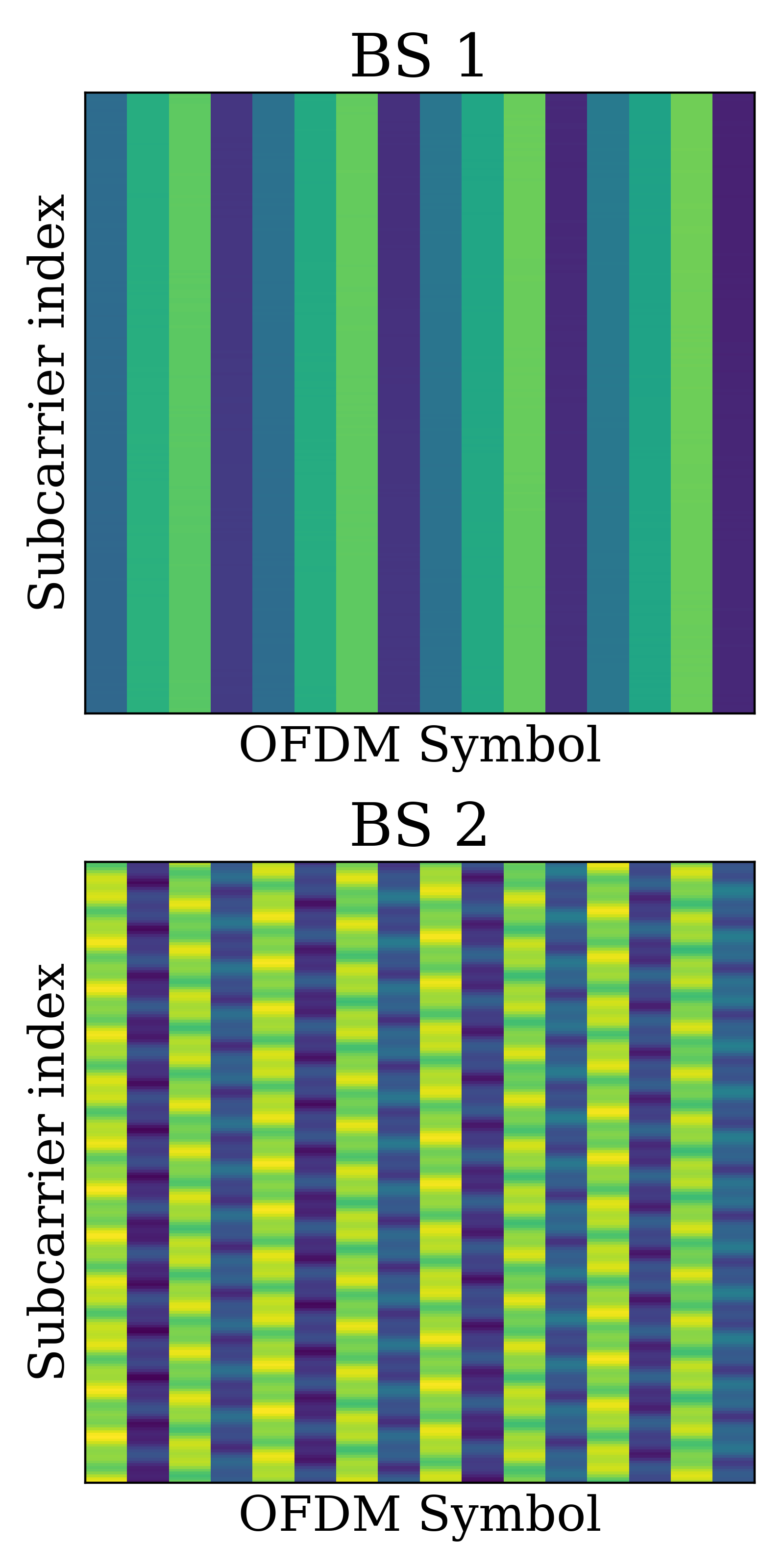}
        \caption{5G NR Positioning}
        \label{fig:positioning}
    \end{subfigure}%
    \begin{subfigure}[t]{0.2\linewidth}
        \centering
        \includegraphics[width=0.8\linewidth,keepaspectratio]{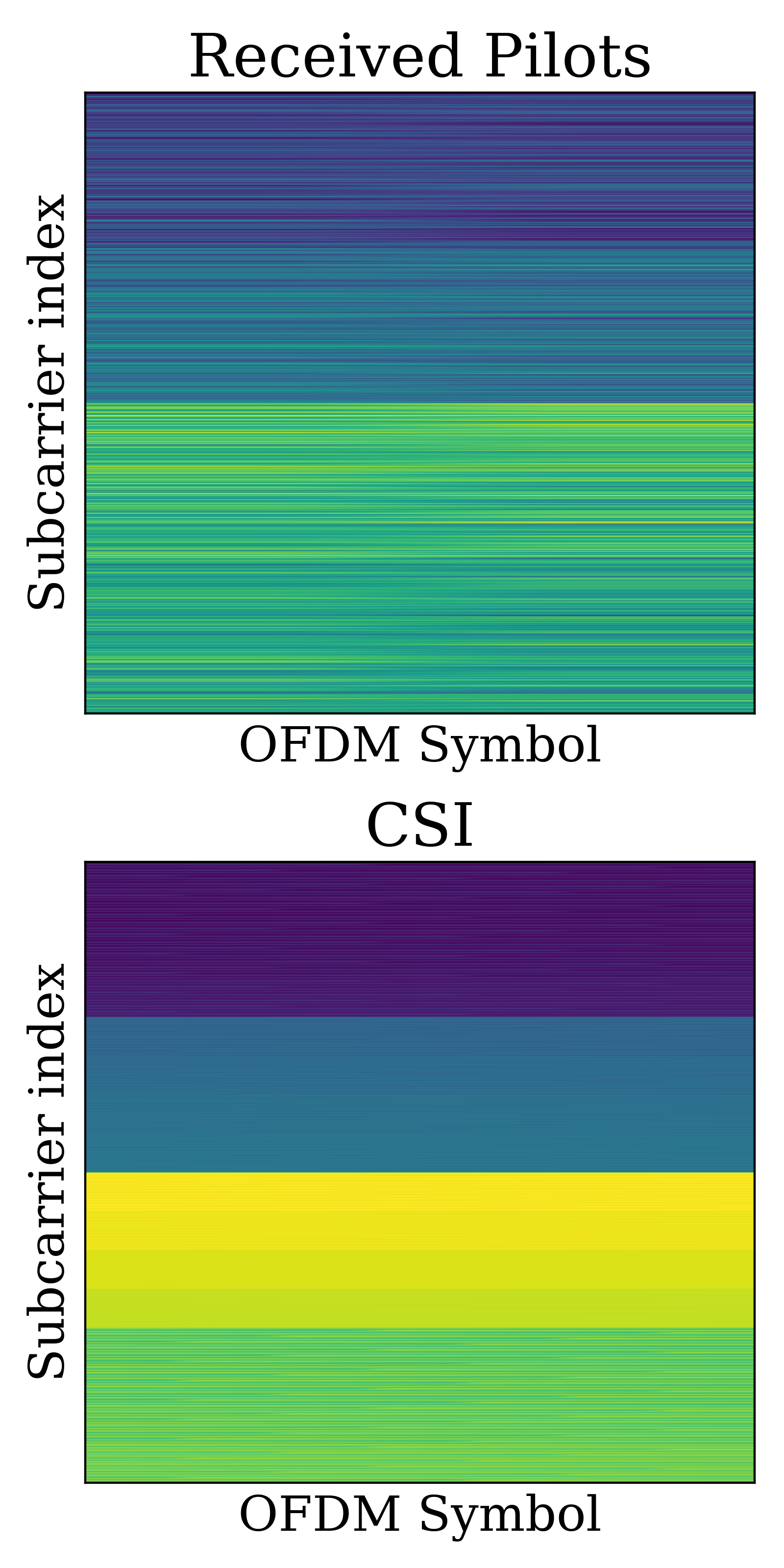}
        \caption{Channel Estimation}
        \label{fig:mimo-ofdm}
    \end{subfigure}
    \captionsetup{justification=centering}
    \caption{Samples from the fine-tuning datasets.}
    \label{fig:finetuning-samples}
\end{figure*}

\section{Related Work}
\label{sec:related_work}

Several studies have highlighted the role of FMs in shaping future 6G networks and their potential applications. In \cite{big_ai_6g, llm_telecom, fm_phy}, a vision is presented for WFMs that can be fine-tuned to perform a wide range of wireless-specific tasks. The adoption of large models, inspired by the large language model (LLM) paradigm, is emphasized in \cite{big_ai_6g, llm_telecom}. These WFMs are envisioned to support network planning, design, deployment, and operation. They are expected to enhance performance through intelligent user access, scheduling, autonomous network management, and adaptive transmission schemes. In \cite{fm_phy}, there is more focus on encouraging the development of WFMs that perform multiple physical‑layer functions.

Adapting LLMs for wireless networks has been investigated in several studies \cite{wireless_llm, llm_multiagent}, with an in-depth overview provided by \cite{llm_survey}. These approaches rely on natural language and existing LLMs to generate algorithms or answer telecom-related questions. They do not directly ingest or represent any raw signals or wireless modalities. In \cite{wireless_llm}, a pipeline is proposed for creating an LLM with wireless domain knowledge through continual pre-training, instruction tuning, prompt engineering, and retrieval-augmented generation. To address limitations in domain-specific knowledge and reasoning, \cite{llm_multiagent} augments LLMs with specialized agents for data retrieval, planning, and self-evaluation. However, these approaches are not in scope, as our focus is on foundation models that are designed to directly process and represent wireless modalities.

Direct modeling of wireless modalities to enable multi-task foundation models has been investigated in \cite{rfm_gc, aboulfotouh2024building6gradiofoundation, wirelessgpt, contrastivefm}. 
The work in \cite{rfm_gc, aboulfotouh2024building6gradiofoundation} presented the first WFM that supports multiple tasks using an SSL approach to extract generalizable representations from RF spectrograms. Results demonstrated effective performance on human activity sensing and spectrogram segmentation with a single shared foundation model.
Wireless channel representation learning has also been recently demonstrated in \cite{wirelessgpt, contrastivefm}. 
Yang et. al \cite{contrastivefm} presented a contrastive learning approach to develop a wireless channel foundation model, and Jiang \cite{wirelessgpt} utilized a masked reconstruction approach similar to \cite{aboulfotouh2024building6gradiofoundation}.
The supported tasks include channel estimation and prediction, line‑of‑sight (LOS) and non‑line‑of‑sight (NLOS) channel identification, and human activity prediction. Both of these approaches center on processing of simulated wireless channel information during pre-training and fine-tuning. 

This paper presents WavesFM, a comprehensive wireless foundation model framework, capable of supporting a wide array of communication, sensing, and localization tasks. Different from prior work, the model is pre-trained solely on real-world data and evaluated on diverse real-world downstream tasks (with the exception of the channel estimation task). The diverse wireless data used during pretraining enables the FM to learn unified, cross‑modal representations. The model is capable of processing both image‑like wireless modalities such as spectrograms and channel state information (CSI), and IQ signals arranged as OFDM resource grids.
Another unique aspect of WavesFM is that it incorporates LoRA to enable efficient fine-tuning for multi-task support.


\section{Testbed and Datasets}
\label{sec:datasets}

We utilize three datasets for pre-training and four datasets for fine-tuning, all from different sources. The data modality is either spectrograms, CSI, or IQ signals collected in various scenarios. We describe them in detail in the following subsections. Samples from the pre-training and fine-tuning datasets are illustrated in Figures \ref{fig:pretraining-samples} and \ref{fig:finetuning-samples}, respectively.

\subsection{Pre-training Datasets}

\textbf{RF Spectrograms (RF-S)}. This dataset consists of spectrograms of various RF signals captured over-the-air. Data was collected for a range of signal types, including LTE, Bluetooth, and WiFi, analog FM, along with additional samples from the ISM band. Each recording is defined by a center frequency in the sub-$6$ GHz band, a sampling frequency between $10$ MHz and $60$ MHz, and a duration that typically averages around $100$ ms. The data collection was conducted in downtown Toronto, Canada, and the dataset has a total of $3,332$ spectrogram samples.

{\textbf{WiFi CSI Samples (WiFi-CSI)}}. This dataset consists of CSI measurements captured using two WiFi routers, one as a transmitter (operating in 802.11n AP mode at $5$ GHz with a $40$ MHz bandwidth) and a three-antenna receiver. 
Experiments were conducted in two indoor environments: a lab testing area, where devices were positioned at the entry to monitor user arrivals and departures, and a cubic office space designed to simulate scenarios of users walking within a confined area. The dataset has a total of $840$ WiFi CSI samples. Although originally gathered for tasks like human authentication and intruder detection \cite{wifi_csi_pretraining}, we employ these CSI samples exclusively for pre-training without utilizing the original task labels.

{\textbf{5G CSI Samples (5G-CSI)}}. This dataset comprises CSI measurements collected with commercial off-the-shelf 5G equipment. The setup includes a baseband unit and a four-channel picocell sub-6GHz remote radio unit that together form a base station, as well as a user terminal (UT) mounted on an autonomous vehicle. Data were gathered in the underground parking lot of an office building, where the UT was positioned at $476$ distinct locations. At each location, repeated uplink measurements were captured using a series of reference symbols, and the corresponding CSI data were logged. The dataset has a total of $476$ samples. Although originally collected for localization \cite{5g-csi}, we use these CSI measurements solely for pre-training, disregarding the original labels.

\textbf{Data Pre-processing}. The pre-processing pipeline for pre-training data is as follows:
1) if the input is a spectrogram, transform it to a logarithmic scale to reduce its sparsity (leave CSI inputs unchanged); 2) normalize the result to the range $[0, 1]$ using dataset-wide statistics; 3) resize the output to $224 \times 224$ pixels using bicubic interpolation; and 4) standardize the output channel-wise using dataset-wide statistics. Normalization and standardization help stabilize the learning process. Additionally, we perform data augmentation to mitigate the size discrepancy between datasets, which ensures balanced performance across modalities.

\subsection{Fine-tuning Datasets}

\textbf{Human Activity Sensing}. This dataset contains CSI measurements for six human activities: running, walking, falling, boxing, arm circling, and floor cleaning \cite{csi_sensing}. Human subjects perform these activities between a pair of Wi-Fi access points, each equipped with three antennas. CSI is measured for each activity, across $114$ subcarriers and $3$ channels (one per antenna) at a $500$ Hz rate, with every recording labeled according to the performed activity. Figure \ref{fig:human_sensing} displays the CSI of the first antenna for samples from the data.

\textbf{RF Signal Classification}. This dataset includes spectrograms representing various signal types, such as WiFi, FM radio, Cellular, Bluetooth, and others, with a total of $20$ classes \cite{Zahid2024}. Samples from the dataset are illustrated in Figure \ref{fig:rf_classification}.

\textbf{5G New Radio (NR) Positioning}. This dataset contains CSI measurements obtained using observations of 5G sounding reference signals exchanged between a user equipment (UE) and four MIMO base stations \cite{5g_positioning}. Each CSI sample is labeled with its real location. The CSI is measured across $192$ subcarriers and $14$ OFDM symbols. Figure \ref{fig:positioning} displays the CSI measurements at an arbitraty UE location for base stations 1 (BS1) and 2 (BS2).

\textbf{MIMO-OFDM Channel Estimation}. The uplink of MIMO-OFDM system is simulated using Sionna \cite{sionna} where the user has a single antenna and the base station has $16$ antennas. We employ the Urban Microcell (UMi) channel model specified by 3GPP. Each transmission comprises $14$ consecutive OFDM symbols, with pilot symbols positioned at indices $2$ and $11$. Channel estimation is conducted at these pilot positions and interpolated for the remaining OFDM symbols that carry data. The system operates at a carrier frequency of $3.5$ GHz, with a subcarrier spacing of $30$ kHz and an average user speed of $3$ m/s. To visualize a sample from the data, we the stack the CSI from each antenna vertically to form a single image as shown in Figure \ref{fig:mimo-ofdm}.

{\textbf{Data Pre-processing}.} For human activity sensing, RF signal classification, and 5G NR positioning, we apply the following pre-processing steps: 1) resize the input to $224 \times 224$ pixels using bicubic interpolation; 2) normalize the pixel values to the range $[0, 1]$; and 3) perform channel-wise standardization using dataset-wide statistics.

For MIMO-OFDM channel estimation, the pre-processing steps are: 1) normalize the reference and received pilots to the range $[-1, 1]$; 2) standardize the data using dataset-wide statistics; and 3) resize the result to $224 \times 224$ pixels using bicubic interpolation. As in pre-training, normalization and standardization help stabilize fine-tuning.

\begin{figure*}[h!]
    \centering
    \begin{subfigure}[t]{\linewidth}
        \centering
        \includegraphics[width=0.9\linewidth, keepaspectratio]{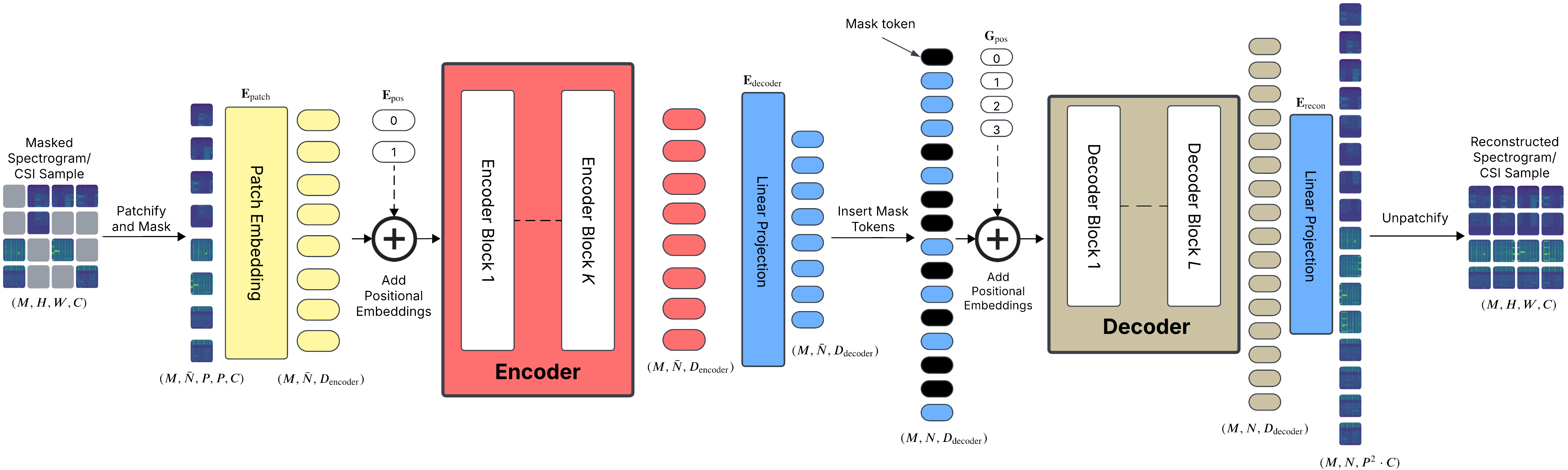}
        \caption{Masked patch prediction with a ViT-based encoder–decoder.}
        \label{fig:pretraining}
    \end{subfigure}
    \begin{subfigure}[t]{0.6\linewidth}
        \centering
        \includegraphics[width=\linewidth, keepaspectratio]{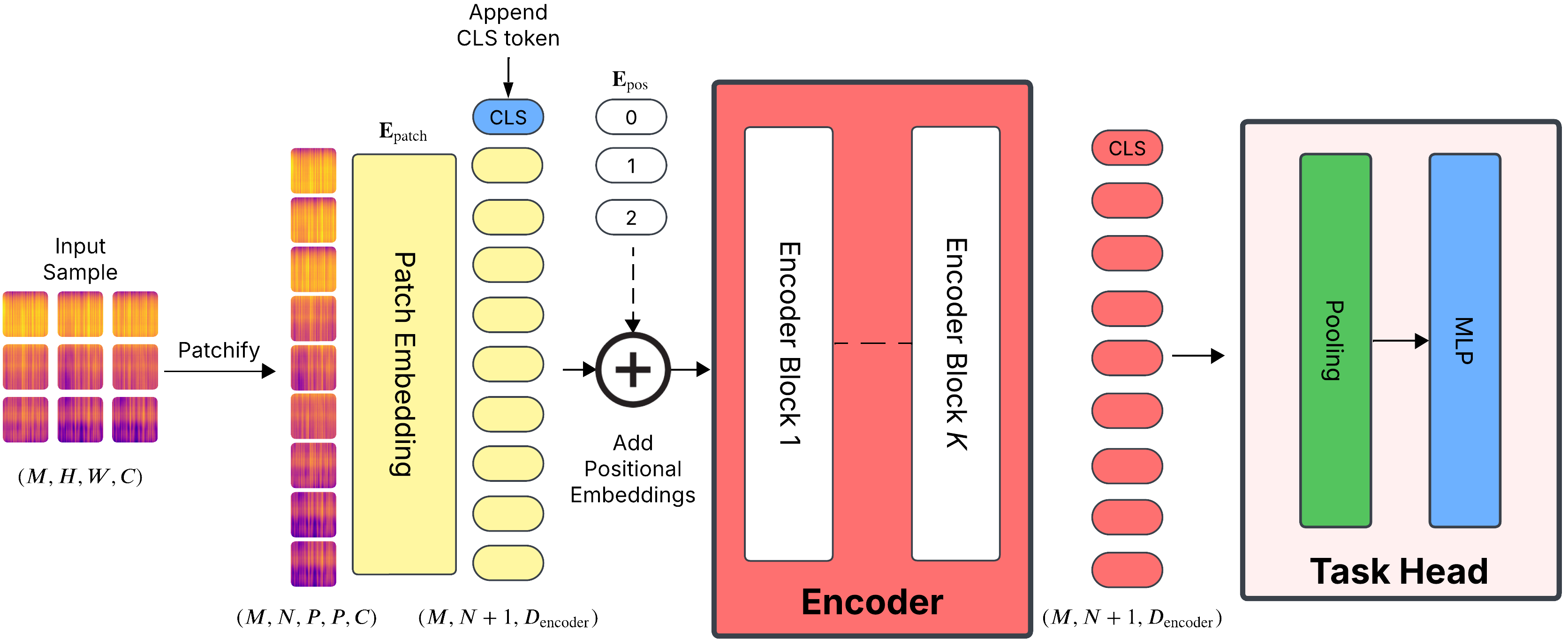}
        \caption{Conventional fine-tuning of the pre-trained model for a downstream task.}
        \label{fig:finetuning}
    \end{subfigure}
    \hspace{2em}
    \begin{subfigure}[t]{0.15\linewidth}
        \centering
        \includegraphics[width=0.8\linewidth, keepaspectratio]{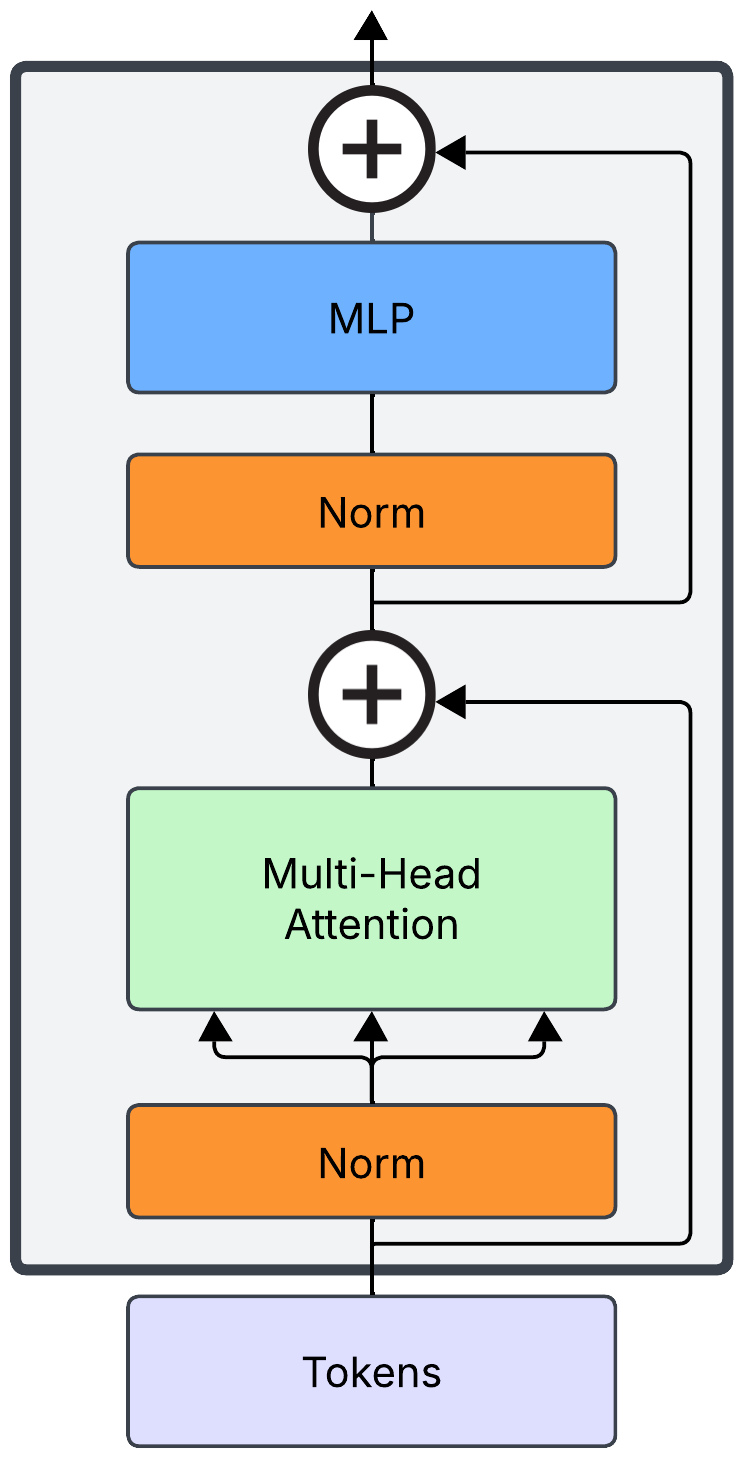}
        \caption{ViT Block.}
        \label{fig:vit}
    \end{subfigure}
    \captionsetup{justification=centering}
    \caption{Overview of the proposed methodology. Figure \ref{fig:pretraining} illustrates masked wireless modeling pre-training, Figure \ref{fig:finetuning} depicts the conventional fine-tuning process and Figure \ref{fig:vit} shows the ViT block internal structure.}
    \label{fig:pretraining-finetuning}
\end{figure*}

\section{WavesFM}
\label{sec:methods}

 In this section, we describe the procedure for building WavesFM illustrated in Figure \ref{fig:rfm_ojcoms}. The approach is divided into two main phases: pre-training and downstream fine-tuning which is further detailed in Figure \ref{fig:pretraining-finetuning}. Figures \ref{fig:pretraining} and \ref{fig:finetuning} illustrate the pre-training and fine-tuning procedures, along with the core components of the model, while Figure \ref{fig:vit} provides an inside look of each ViT block. We train the model in a self-supervised way, to learn a robust and general representation of the data without the relying on labels. The resulting WFM provides a versatile backbone that can be used across multiple wireless sensing, communication and localization tasks.

In the downstream fine-tuning stage, we adapt the WFM to specific tasks using fine-tuning datasets. Here, the model is shared as extensively as possible between tasks to promote multi-task capabilities. We present two fine-tuning approaches: conventional fine-tuning, which selectively adapts certain blocks of the ViT and share the remaining with other tasks, and LoRA, a parameter-efficient method that freezes the whole WFM and fine-tunes only a minimal set of additional task-specific parameters.

\subsection{Masked Autoencoder Pre-training}

We utilize a masked wireless modeling (MWM) technique with the ViT for pre-training. 
In this approach, each input, whether a spectrogram or a CSI sample, is divided into patches, and a random subset of these patches is selected to be masked. The model is then tasked with reconstructing the masked patches from the visible ones. 
By analyzing the visible patches, the model infers the likely content of the masked regions. Successfully reconstructing the full sample from only partial observations indicates that the model likely learned an internal representation of the underlying data distribution.
While this approach resembles a traditional auto-encoder, a key difference is that the model is trained to reconstruct the masked patches only, rather than the full set. 

To reduce redundancy and make reconstruction more challenging, we employ high masking ratios (e.g., $75\%$). This forces the model to rely less on extrapolation from visible patches, and more on learning useful and general features of the input. This approach offers several advantages: masking a large portion of the input and only processing the sampled patches makes pre-training more efficient. There is no need for the data to be labeled, because the data itself provides the self-supervision needed for learning.

We employ the ViT masked autoencoder architecture proposed in \cite{mae}. The design consists of an encoder that processes the visible patches to produce feature tokens, and a decoder that reconstructs the original input by processing the encoder output and mask tokens. Mask tokens are learnable embeddings positioned at the original locations of the masked patches, signaling to the decoder where reconstruction is required.
The encoder is designed to have a higher capacity than the decoder, so it performs the majority of the computation. Hence, the encoder can function independently as a feature extractor, and the decoder is discarded after pre-training. In the following, we provide additional details on the architecture. The end-to-end pre-training approach is illustrated in Figure \ref{fig:pretraining} and detailed in Algorithm \ref{alg:pretraining}.

A single ViT block is illustrated in Figure \ref{fig:vit}, which consists of a multi-headed self-attention (MSA) module preceded by layer normalization, followed by a multilayer perceptron (MLP) block that is also preceded by layer normalization. The model takes an image-like tensor of shape $(H, W, C)$ as an input, which may represent spectrogram, CSI or another modality. The input is mapped to a sequence of flattened patches $\mathbf{X} = [\bm{x}_{1}, \cdots, \bm{x}_{N}] \in \mathbb{R}^{N \times P^2 \cdot C}$, $\bm{x}_i \in \mathbb{R}^{P^2 \cdot C}$ where $(H, W)$ is the resolution of the original image, $C$ is the number of channels, $(P, P)$ is the resolution of each patch and $N = HW/P^2$ is the number of resulting patches. Both the encoder and decoder consist of a series of ViT blocks, but they differ in their embedding dimensions, denoted as $D_\text{encoder}$ and $D_\text{decoder}$, respectively. 

{\textbf{Masking}.} During pre-training, we mask a fraction $\gamma \in (0, 1)$ of the patches by removing them from the input. The patches to be removed are chosen using a uniform distribution. To avoid center bias, we first shuffle the patches, then randomly select a subset for removal, and finally restore the original order of the remaining patches. After masking, the masked patches are then flattened and collected into the sequence
\begin{equation}
    \label{eq:masked_patches}
    \mathbf{X}_p = [\mathbf{x}_{p,1}, \dots, \mathbf{x}_{p,\tilde N}] \in \mathbb{R}^{\tilde N \times P^2 C}
\end{equation}
where $\mathbf{x}_{p,i}\in\mathbb{R}^{P^2\cdot C}$ is the $i$‑th masked patch and $\tilde N = \lfloor \gamma N\rfloor$ is the number of masked patches.  

The remaining (visible) patches are also flattened and collected into the sequence
\begin{equation}
    \label{eq:visible_patches}
    \bar{\mathbf{X}}_p = [\bar{\mathbf{x}}_{p,1}, \dots, \bar{\mathbf{x}}_{p,\bar N}] \in \mathbb{R}^{\bar N \times P^2 C},
\end{equation}
where $\bar{\mathbf{x}}_{p,i}\in\mathbb{R}^{P^2\cdot C}$ is the $i$‑th visible patch and $\bar N = N - \lfloor \gamma N\rfloor$ is the number of visible patches.

{\textbf{Patch Embedding}}. Each visible patch is projected to match the encoder embedding dimension using a learnable linear projection matrix $\mathbf{E}_{\text{patch}} \in \mathbb{R}^{D_{\text{encoder}} \times P^2 \cdot C}$. Next, sine-cosine positional embeddings are added to form a token. The resulting sequence $\mathbf{Z}_0 \in \mathbb{R}^{\bar{N} \times D_{\text{encoder}}}$ can be written as:
\begin{equation}
    \label{eq:patch_embedding}
    \mathbf{Z}_0 = [\bar{\bm{x}}_{p,1}\mathbf{E}_{\text{patch}}, \cdots, \bar{\bm{x}}_{p,\bar{N}}\mathbf{E}_{\text{patch}}] + \mathbf{E}_{\text{pos}}
\end{equation}
where $\mathbf{E}_{\text{pos}} \in \mathbb{R}^{\bar{N}\times D_{\text{encoder}}}$ denotes the encoder's positional embedding matrix.

Positional embeddings are used to indicate token ordering in Transformer-based architectures, which otherwise have no inherent mechanism for representing positional information. 
Once the patch embedding projection is applied, each resulting token is associated with a specific location on a two-dimensional grid by encoding its corresponding coordinates.

Given an input sequence in $\mathbb{R}^{N \times D}$, the embedding dimension $D$ is split equally: the first half encodes the row coordinate, and the second half encodes the column coordinate. The process to generate the positional embeddings is as follows: 1) Create a standard 1D sine-cosine embedding for the row in $\mathbb{R}^{D/2}$; 2) Create a standard 1D sine-cosine embedding for the column in $\mathbb{R}^{D/2}$; and 3) Concatenate these two vectors to form the complete row-column embedding in $\mathbb{R}^D$.
Formally, the 1D sine-cosine embedding for a given position $i$ is given by:
\begin{align}
    &[\sin(i\cdot \omega_0), \cdots, \sin(i\cdot \omega_{\frac{M}{2}-1}),\nonumber\\ 
    &\cos(i\cdot \omega_0), \cdots, \cos(i\cdot \omega_{\frac{M}{2}-1})] \quad \in \mathbb{R}^M
\end{align}
where $\omega_k = 10000^{-\frac{2k}{M}}$ and $M=D/2$ following the original Transformer formulation \cite{attention}. The resulting positional embeddings are then added to the sequence of tokens as expressed in equation \eqref{eq:patch_embedding} and illustrated in Figure \ref{fig:pretraining}. 

\begin{algorithm}[h!]
    \DontPrintSemicolon
    \SetKwInOut{Input}{Input}
    \SetKwInOut{Output}{Output}
    \SetKwInOut{Initialization}{Initialization}
    \Input{dataset $\mathcal{D}$, ViT masked autoencoder $\mathcal{M}$, mask ratio $\gamma$}
    \Output{foundation model $\mathcal{F}$}
    \Initialization{$\mathcal{F} \gets $ ViT encoder from from $\mathcal{M}$, $P \gets$ patch size. $K \gets$ number of encoder blocks, $L \gets$ number of decoder blocks.}
    \nl{}\Repeat{\normalfont convergence is reached.}{
    \nl\For{\normalfont each batch $\mathcal{B}$ \textbf{in} $\mathcal{D}$}{
    \nl{}Initialize the loss value: $\mathcal{L}$ $\gets 0$ \\
    \nl{}\For{\normalfont each sample $\mathbf{X}$ \textbf{in} $\mathcal{B}$}{
    \nl{}Map $\mathbf{X}$ into a sequence of $P \times P$ flattened patches: $\left[\bm{x}_{1}, \cdots, \bm{x}_{N}\right]$.\\
    \nl{}Sample $\bar{N} = N - \lfloor\gamma N\rfloor$ from the sequence: $\left[\bar{\bm{x}}_{p,1}, \cdots, \bar{\bm{x}}_{p,\bar{N}}\right]$.\\
    \nl{}Perform patch embedding to create the encoder input $\mathbf{Z}_0$ as in equation \eqref{eq:patch_embedding}.\\
    \nl{}Perform the encoder forward pass:\\
    \nl{}\For{$k \gets 1$ \KwTo $K$}{
    $\bar{\mathbf{Z}}_{k} \gets \text{MSA}_{k}(\text{LN}_{k}(\mathbf{Z}_{k-1})) + \mathbf{Z}_{k - 1}$\\
    $\mathbf{Z}_k \gets \text{MLP}_{k}(\bar{\text{LN}}_{k}(\bar{\mathbf{Z}}_k))) + \bar{\mathbf{Z}}_k$\\}
    \nl{}\textbf{end}\\
    \nl{}Perform decoder embedding on $\mathbf{Z}_K$ and insert mask tokens to create the decoder input $\mathbf{Y}_0$ as in equation \eqref{eq:decoder_embedding}.\\
    \nl{}Perform the decoder forward pass:\\
    \nl{}\For{$l \gets 1$ \KwTo $L$}{
    $\bar{\mathbf{Y}}_{l} \gets \text{MSA}_{l}(\text{LN}_{l}(\mathbf{Y}_{l-1})) + \mathbf{Y}_{l - 1}$\\
    $\mathbf{Y}_l \gets \text{MLP}_{l}(\bar{\text{LN}}_{l}(\bar{\mathbf{Y}}_l))) + \bar{\mathbf{Y}}_l$\\}
    \nl{}\textbf{end}\\
    \nl{}Reconstruct patches from the decoder output using equations \eqref{eq:recon_patches_all} and \eqref{eq:recon_patches} to get the reconstructed patches $\hat{\mathbf{X}}_p$.\\
    \nl{}Accumulate the MWM loss defined in equation \eqref{eq:loss_masked} using $\hat{\mathbf{X}}_p$ and masked patches $\mathbf{X}_p$ in equation \eqref{eq:masked_patches}: $\mathcal{L} \mathrel{+}= \mathcal{L}_{\text{MWM}}(\mathbf{X}_p, \hat{\mathbf{X}}_p)$}
    \nl{}{\normalfont\textbf{end}}\\
    \nl{}Update parameters of the model $\mathcal{M}$ using the average loss $\mathcal{L}/|\mathcal{B}|$ via backpropagation.\\}
    \nl{\normalfont\textbf{end}}}
    \vspace{3pt}
    {\noindent Note that processing each batch is done in parallel, here it is shown as a loop for convenience.}
    \caption{Masked Wireless Modeling with ViT}
    \label{alg:pretraining} 
\end{algorithm}

{\textbf{ViT Block}.} The output of the $l$-th ViT block is computed using a residual architecture that combines MSA with an MLP block, with each block preceded by layer normalization, as shown in Figure \ref{fig:vit}. For a generic sequence of tokens $\mathbf{Z}_l \in \mathbb{R}^{N\times D}$, this is expressed as:
\begin{align}
    \label{eq:transformer_block_msa}
    \bar{\mathbf{Z}}_{l} &= \text{MSA}_{l}(\text{LN}_{l}(\mathbf{Z}_{l-1})) + \mathbf{Z}_{l - 1} \\
    \label{eq:transformer_block_mlp}
    \mathbf{Z}_l &= \text{MLP}_{l}(\bar{\text{LN}}_{l}(\bar{\mathbf{Z}}_l))) + \bar{\mathbf{Z}}_l
\end{align}
where LN and $\bar{\text{LN}}$ denote layer-normalization applied before MSA and MLP, respectively and $\mathbf{Z}_l$ is the output of the $l$-th block. The layer-normalization operation is described in \cite{layernormalization}.

For an input sequence $\mathbf{Z} \in \mathbb{R}^{N\times D}$, the self-attention (SA) mechanism projects $\mathbf{Z}$ using a learnable projection matrix $\mathbf{U}_{qkv} \in \mathbb{R}^{D\times 3D}$ as follows:
\begin{align}
    [\mathbf{Q}, \mathbf{K}, 
    \mathbf{V}] &= \mathbf{Z}\mathbf{U}_{qkv} \in \mathbb{R}^{N\times3D}.
\end{align}
We then partition the resulting matrix into the query $\mathbf{Q}$, key $\mathbf{K}$, and value $\mathbf{V}$ matrices, each of dimension $\mathbb{R}^{N\times D}$.
Next, we compute the attention scores by taking the scaled dot product between the queries, and keys. The softmax function is then applied row-wise to normalize the attention scores, ensuring that each row forms a probability distribution over the input tokens, and the result is multiplied by the values which is expressed as: 
\begin{equation}
    \mathbf{A} = \text{softmax} \left(\mathbf{Q}\mathbf{K}^T /\sqrt{D}\right) \mathbf{V} \in \mathbb{R}^{N\times D}
\end{equation}
MSA is an extension of SA, where we run $k$ SA operations, and their outputs are concatenated, then projected using a learnable projection matrix $\mathbf{U}_{\text{msa}} \in \mathbb{R}^{k\cdot D \times D}$ as follows:
\begin{equation}
    \text{MSA} = [\mathbf{A}_1, \cdots, \mathbf{A}_k] \mathbf{U}_{\text{msa}} \in \mathbb{R}^{N \times D}
\end{equation}
where $\mathbf{A}_i \in \mathbb{R}^{N\times D}$ is the attention score for the $i$-th head.

The MLP consists of two fully connected linear layers with a GELU activation in between. The first layer projects the input into a hidden dimension of $D_h$, and the second layer maps this hidden representation back to the transformer’s latent dimension $D$. 

After performing patch embedding and adding positional embeddings as in equation \eqref{eq:patch_embedding}, the encoder processes the sequence through a series of $K$ ViT blocks using equations \eqref{eq:transformer_block_msa} and \eqref{eq:transformer_block_mlp}, which yields feature tokens $\mathbf{Z}_K=\left[\bm{z}_{K,1},\cdots,\bm{z}_{K,\bar{N}}\right] \in \mathbb{R}^{\bar{N} \times D_{\text{encoder}}}$. These tokens are then projected to the decoder’s embedding space using a learnable linear projection matrix $\mathbf{E}_{\text{decoder}} \in \mathbb{R}^{D_{\text{encoder}} \times D_{\text{decoder}}}$. 

\noindent Subsequently, a mask token $\bm{y}_{\text{mask}} \in \mathbb{R}^{D_{\text{decoder}}}$ is inserted in the positions of masked patches, and positional embeddings are added. If we assume a decoder positional embedding matrix $\mathbf{G}_{\text{pos}} = \left[\bm{g}_{\text{pos},1}, \cdots, \bm{g}_{\text{pos},N}\right] \in \mathbb{R}^{N \times D_{\text{decoder}}}$, a token in the decoder input $\mathbf{Y}_0 \in \mathbb{R}^{N\times D_{\text{decoder}}}$ is expressed as:
\begin{equation}
    \label{eq:decoder_embedding}
    \bm{y}_{0,i} = 
    \begin{cases}
        \bm{z}_{K,i}\mathbf{E}_{\text{decoder}} + \bm{g}_{\text{pos},i}, & \text{if position } i \text{ is unmasked},\\[1em]
        \bm{y}_{\text{mask}} + \bm{g}_{\text{pos},i}, & \text{if position } i \text{ is masked}.
    \end{cases}
\end{equation}
The decoder processes the sequence through its series of $L$ ViT blocks whose output is computed according to equations \eqref{eq:transformer_block_msa} and \eqref{eq:transformer_block_mlp} resulting in a sequence $\mathbf{Y}_K = \left[\bm{y}_{K,1}, \cdots, \bm{y}_{K,N} \right] \in \mathbb{R} ^ {N\times D_{\text{decoder}}}$. The output of the decoder is mapped to the reconstructed patches using a learnable projection matrix $\mathbf{E}_{\text{recon}} \in \mathbb{R}^{D_{\text{decoder}} \times P^2\cdot C}$, which is expressed as:
\begin{equation}
    \label{eq:recon_patches_all}
    \hat{\mathbf{X}} = \left[\bm{y}_{K,1} \mathbf{E}_{\text{recon}}, \cdots, \bm{y}_{K,N} \mathbf{E}_{\text{recon}} \right] \in \mathbb{R}^{N \times P^2\cdot C}
\end{equation}
By only keeping the reconstruction of the originally masked patches, we get:
\begin{equation}
    \label{eq:recon_patches}
    \hat{\mathbf{X}}_p = [\hat{\bm{x}}_{p,1}, \cdots, \hat{\bm{x}}_{p,\tilde{N}}] \in \mathbb{R}^{\tilde{N}\times P^2 \cdot C}
\end{equation}
 \textbf{Objective.} We train the model end-to-end to reconstruct the masked patches from the input patches. The loss function $\mathcal{L}_{\text{MWM}}$ of the MWM task can be written in terms of the masked patches and their reconstruction from equations \eqref{eq:masked_patches} and \eqref{eq:recon_patches}, respectively as:
 \begin{equation}
    \label{eq:loss_masked}
     \mathcal{L}_\text{MWM} = \dfrac{1}{M\tilde{N}} \sum_{m=1}^{M} \sum_{i=1}^{\tilde{N}} \|\bm{x}_{p,i}^{(m)} - \hat{\bm{x}}_{p,i}^{(m)} \|^2
 \end{equation}
where $M$ is the batch size and $m$ is the sample index. Note that the loss is computed exclusively for the masked patches, while the visible patches (originally inputted to the model) are omitted from the reconstruction loss. The encoder of the ViT masked autoencoder serves as our WFM which can be fine-tuned for downstream tasks.
\subsection{Downstream Fine-tuning}

\begin{algorithm}[h!]
    \DontPrintSemicolon
    \SetKwInOut{Input}{Input}
    \SetKwInOut{Output}{Output}
    \SetKwInOut{Initialization}{Initialization}
    \Input{dataset $\mathcal{D}$, foundation model $\mathcal{F}$, task head $\mathcal{H}$, loss function $\mathcal{L}_{\text{task}}$, frozen blocks $S$, $f_{\text{LoRA}} \in \{\text{True}, \text{False}\}$ (enable LoRA), low-rank dimension $R$, scaling factor $\alpha$}
    \Output{fine-tuned model $\mathcal{M}$}
    \Initialization{$P \gets$ patch size, $K \gets$ number of encoder blocks.}    
    \nl{}Attach task head to ViT encoder: $\mathcal{M} \gets \mathcal{F} \cup \mathcal{H}$.\\
    \nl{}\If{$f_{\normalfont\text{LoRA}}$}{
        Attach LoRA query and value adapters to $\mathcal{M}$ with rank $R$ and scaling factor $\alpha$\\
        Freeze all blocks from $\mathcal{B}$\\
    }
    \nl{}\Else{
        Freeze $S$ blocks from $\mathcal{B}$\\}
    \textbf{end}\\
    \nl{}\Repeat{\normalfont convergence or stopping condition is met.}{
    \nl\For{\normalfont each batch $\mathcal{B}$ \textbf{in} $\mathcal{D}$}{
    \nl{}Initialize the loss value $\mathcal{L}\gets 0$ \\
    \nl{}\For{\normalfont each sample $\mathbf{X}$, $\mathbf{Y}$ \textbf{in} $\mathcal{B}$}{{
        \nl{}Map the image $\mathbf{X}$ into a sequence of $P \times P$ flattened patches: $\left[\bm{x}_{p,1}, \cdots, \bm{x}_{p,N}\right]$.\\
        \nl{}Perform patch embedding to create the encoder input $\mathbf{Z}_0$ and append CLS token as in equation \eqref{eq:patch_embedding_two}.\\
        \nl{}Perform the encoder forward pass:\\
        \For{$k \gets 1$ \KwTo $K$}{
        $\bar{\mathbf{Z}}_{k} \gets \text{MSA}_{k}(\text{LN}_{k}(\mathbf{Z}_{k-1})) + \mathbf{Z}_{k - 1}$\\
        $\mathbf{Z}_k \gets \text{MLP}_{k}(\bar{\text{LN}}_{k}(\bar{\mathbf{Z}}_k))) + \bar{\mathbf{Z}}_k$\\}
        \textbf{end}\\
        \nl{}Compute the task head prediction using the encoder output: $\hat{\mathbf{Y}} = \mathcal{H}(\mathbf{Z}_K)$.\\
        \nl{}Accumulate the task-specific loss: $\mathcal{L}\mathrel{+}=\mathcal{L}_{\text{task}}(\mathbf{Y}, \hat{\mathbf{Y}})$.\\}}
        \nl{}{\textbf{end}\\}
        \nl{}Update parameters of the model $\mathcal{M}$ using the average loss $\mathcal{L}/|\mathcal{B}|$ via backpropagation.\\}
        \nl{}{\textbf{end}\\}
    }
    \vspace{3pt}
    {\noindent Note that processing each batch is done in parallel, here it is shown as a loop for convenience.}
    \caption{General Fine-tuning}
    \label{alg:finetune}
\end{algorithm}

Fine-tuning the WFM for a downstream task involves appending a task-specific head on top of the ViT encoder, as illustrated in Figure \ref{fig:finetuning}. The selection of the task head depends on the downstream task. For closely related classification and regression tasks, a simple linear layer with the appropriate number of outputs is typically sufficient. The more the downstream task or data distribution deviate from the pre-training domain, fine-tuning a subset of the ViT blocks becomes more necessary. The downstream fine-tuning is detailed in Algorithm \ref{alg:finetune}. 

In the fine-tuning setup, a class token [CLS] is prepend to the input sequence and is commonly used as the feature token for subsequent processing by the task head. The strategy for selecting the feature set from the encoder output is known as pooling, with options such as ‘token’ (which utilizes only the CLS token) or ‘avg’ (which averages all tokens). Note that during fine-tuning, no masking is applied; the full sequence is provided as input to the model. Otherwise, the processing steps of the model are largely the same. 

For a sequence of flattened patches  $\mathbf{X}_p = [\bm{x}_{p,1}, \cdots, \bm{x}_{p,N}] \in \mathbb{R}^{N \times P^2 \cdot C}$, the input $\mathbf{Z}_0 \in \mathbb{R}^{(N + 1) \times D_{\text{encoder}}}$ to the encoder is written as:
\begin{equation}
    \label{eq:patch_embedding_two}
    \mathbf{Z}_0 = [\bm{x}_{\text{CLS}}, \bm{x}_{p,1}\mathbf{E}_{\text{patch}}, \cdots, \bm{x}_{p,N}\mathbf{E} _{\text{patch}}] + \bar{\mathbf{E}}_{\text{pos}}
\end{equation}
where $\bm{x}_{\text{CLS}} \in \mathbb{R}^{D_{\text{encoder}}}$ is the class token, and $\bar{\mathbf{E}}_{\text{pos}} \in \mathbb{R}^{(N+1)\cdot D_{\text{encoder}}}$ is a positional embedding matrix with $N + 1$ tokens.

One of the more sophisticated fine-tuning methods is Low-Rank Adaptation (LoRA), which is a parameter-efficient fine-tuning approach \cite{lora}. LoRA operates by freezing the pre-trained model and introducing low-rank trainable matrices into the attention mechanism of the ViT blocks. 
In the standard self-attention mechanism, the input is projected into query $\mathbf{Q}$, key $\mathbf{K}$, and value $\mathbf{V}$ matrices using learned projection matrices. LoRA introduces additional low-rank matrices to the query and value projections referred to as \textit{adapters}, while leaving the key projection unchanged. Each LoRA adapter consists of two trainable matrices, $\mathbf{A}$ and $\mathbf{B}$, which approximate updates to the model’s parameters in a low-dimensional space.

For an input sequence $\mathbf{Z} \in \mathbb{R}^{N\times D}$, instead of updating the full weight matrix of the query $\mathbf{Q}$ and value $\mathbf{V}$ projections, LoRA introduces the transformation:
\begin{align}
    \mathbf{Q_{\text{LoRA}}} &= \mathbf{Q} + \alpha \mathbf{Z} \mathbf{A}_q \mathbf{B}_q  \in \mathbb{R}^{N\times D}\\
    \mathbf{V_{\text{LoRA}}} &= \mathbf{V} + \alpha \mathbf{Z} \mathbf{A}_v \mathbf{B}_v \in \mathbb{R}^{N\times D}
\end{align}
where $\mathbf{A}_q, \mathbf{A}_v \in \mathbb{R}^{D \times R}$ and $\mathbf{B}_q, \mathbf{B}_v\in \mathbb{R}^{R \times D}$, with $R$ being the rank of the low-rank adaptation and $D$ is the ViT embedding dimension. $R$ controls the number of additional parameters introduced, the higher $R$ is, the larger the low-dimensional subspace and the more parameters are added. $\alpha$ determines the strength of the adaptation. 
The output of the LoRA-enhanced self-attention layer is computed by adding these low-rank modifications to the original query and value projections.

This significantly reduces the number of trainable parameters required for downstream tasks while keeping the pre-trained model itself frozen. As a result, the WFM can be easily shared across multiple downstream tasks, with only the task-specific LoRA adapters and the task-head being fine-tuned. As with conventional fine-tuning, the output of the encoder is then used for subsequent processing by the task head.

\subsection{Task Fine-tuning Configuration}

For the downstream tasks, we consider two classification tasks, human activity sensing and RF signal classification, and two regression tasks, 5G NR positioning and MIMO-OFDM channel estimation. The rationale behind selecting these tasks is to encompass a broad spectrum of domains: two distinct sensing tasks, a communication task, and a localization task.

\textbf{Human Activity Sensing.} The task is to classify CSI measurements into one of six distinct human activity classes. We attach a $6$-neuron linear layer as a task head to the WFM. The model outputs a softmax probability vector, and the loss function is the label smoothing cross-entropy $\mathcal{L}_{\text{SCE}}$, defined as:
\begin{equation}
    \label{eq:loss_sce}
    \mathcal{L}_{\text{SCE}} = -\dfrac{1}{N} \sum_{n=1}^N \sum_{i=1}^{C} \left(y_{i}^{(n)} (1 - \theta) + \dfrac{\theta}{C}\right) \log\left(\hat{y}_{i}^{(n)}\right)
\end{equation}
where $N$ is the batch size, $C = 6$ is the number of classes, $y_i^{(n)} \in [0, 1]$ is the true label for class $i$ (either $0$ or $1$ for sample $n$), $\hat{y}_i^{(n)} \in [0, 1]$ is the model’s predicted probability for class $i$, and $\theta \in (0, 1)$ is the smoothing factor. Unlike traditional cross-entropy, label smoothing distributes a small probability to incorrect labels, preventing the model from becoming overly confident which enhances generalization. The degree of smoothing is determined by $\theta$.

\textbf{RF Signal Classification.} The task is to classify spectrograms into one of 20 distinct signal types. We attach a $20$-neuron linear layer as a task head to the WFM. The model outputs a softmax probability vector, and the loss function is the weighted cross-entropy $\mathcal{L}_{\text{WCE}}$ defined as:
\begin{equation}
    \label{eq:loss_wce}
    \mathcal{L}_{\text{WCE}} = -\dfrac{1}{N} \sum_{n=1}^N \sum_{i=1}^{C} \beta_i y_{i}^{(n)} \log\left(\hat{y}_{i}^{(n)}\right)
\end{equation}
where $N$ is the batch size, $C = 20$ is the number of classes, $y_i^{(n)} \in [0, 1]$ is the true label for class $i$ (either $0$ or $1$ for sample $n$), $\hat{y}_i^{(n)} \in [0, 1]$ is the model’s predicted probability for class $i$, and $\beta_i$ is the weight assigned to class $i$. Unlike traditional cross-entropy, weighted cross entropy helps mitigate class imbalance which is the case with this dataset.

\textbf{5G NR Positioning.} The task is to predict a UE position given CSI observations from $4$ base stations as an input. We attach a $3$-neuron linear layer as a task head to the WFM. The model outputs a position vector $\hat{\bm{r}}^{(n)} \in \mathbb{R}^3$ where $n$ is the sample index, and the loss function is the traditional MSE loss $\mathcal{L}_{\text{MSE}}$ defined as:
\begin{equation}
    \label{eq:loss_mse}
    \mathcal{L}_{\text{MSE}} =  \dfrac{1}{N}\sum_{n=1}^N \| \hat{\bm{r}}^{(n)} - \bm{r}^{(n)} \|^2 .
\end{equation}
where $\bm{r}^{(n)} \in \mathbb{R}^3$ is the original UE position.

\textbf{MIMO-OFDM Channel Estimation.} The task to estimate the channel between a user and $16$-antenna base station using uplink reference OFDM symbols. Every $14$ OFDM symbols, pilots are transmitted on symbols $2$ and $11$. We attach a single ViT block as a task head to the WFM. The model utilizes the received IQ reference symbols and the known transmitted IQ pilots, arranged in an OFDM resource grid, to estimate the channel. To ensure robust performance across a range of signal-to-noise ratio (SNR) values, we use a weighted MSE loss. The key idea is to compensate for larger errors at lower SNRs by assigning lower weights and to boost the contribution of higher SNRs with higher weights, so that the weighted errors become balanced. Specifically, we utilize the following weighting function:
\begin{equation}
    \label{eq:snr_weights}
    w(\text{SNR}) = 10.6\ e^{0.226\times\text{SNR}} - 0.764 \in \mathbb{R}
\end{equation}
which yields a weight at $20$ dB that is 100 times larger than the weight at $-10$ dB. This function is fitted based on the channel estimation performance of the least-squares (LS) estimator on our MIMO-OFDM channel estimation dataset. In this calibration, the error at $-20$ dB, treated as the maximum error, is assigned a weight of $1$, and the weights for SNRs in the range $(-20,10]$ dB are scaled so that their average magnitudes are equal to that of the error observed at $-20$ dB.
The SNR-weighted MSE loss over $N$ samples is defined as
\begin{equation}
    \label{eq:loss_snr_mse}
    \mathcal{L}_{\text{SNR-MSE}} = \frac{1}{N} \sum_{n=1}^{N} w\left(\text{SNR}^{(n)}\right) \cdot \left\| \bm{h}^{(n)} - \hat{\bm{h}}^{(n)} \right\|^2
\end{equation}
where $\bm{h}^{(n)}$ is the flattened ground truth CSI for the $n$-th sample, $\hat{\bm{h}}^{(n)}$ is the corresponding CSI estimate, and $\text{SNR}^{(n)}$ denotes the sample's signal-to-noise ratio.

\section{Results and Discussion}
\label{sec:results}

In this section, we present the performance evaluation of WavesFM across all downstream tasks. As a baseline, we use supervised learning (SL) on each dataset (training from scratch with a model of identical size and capacity to the WFM), which we refer to as ViT‑SL.

\begin{figure}[h!]
    \centering
    \includegraphics[width=0.9\linewidth]{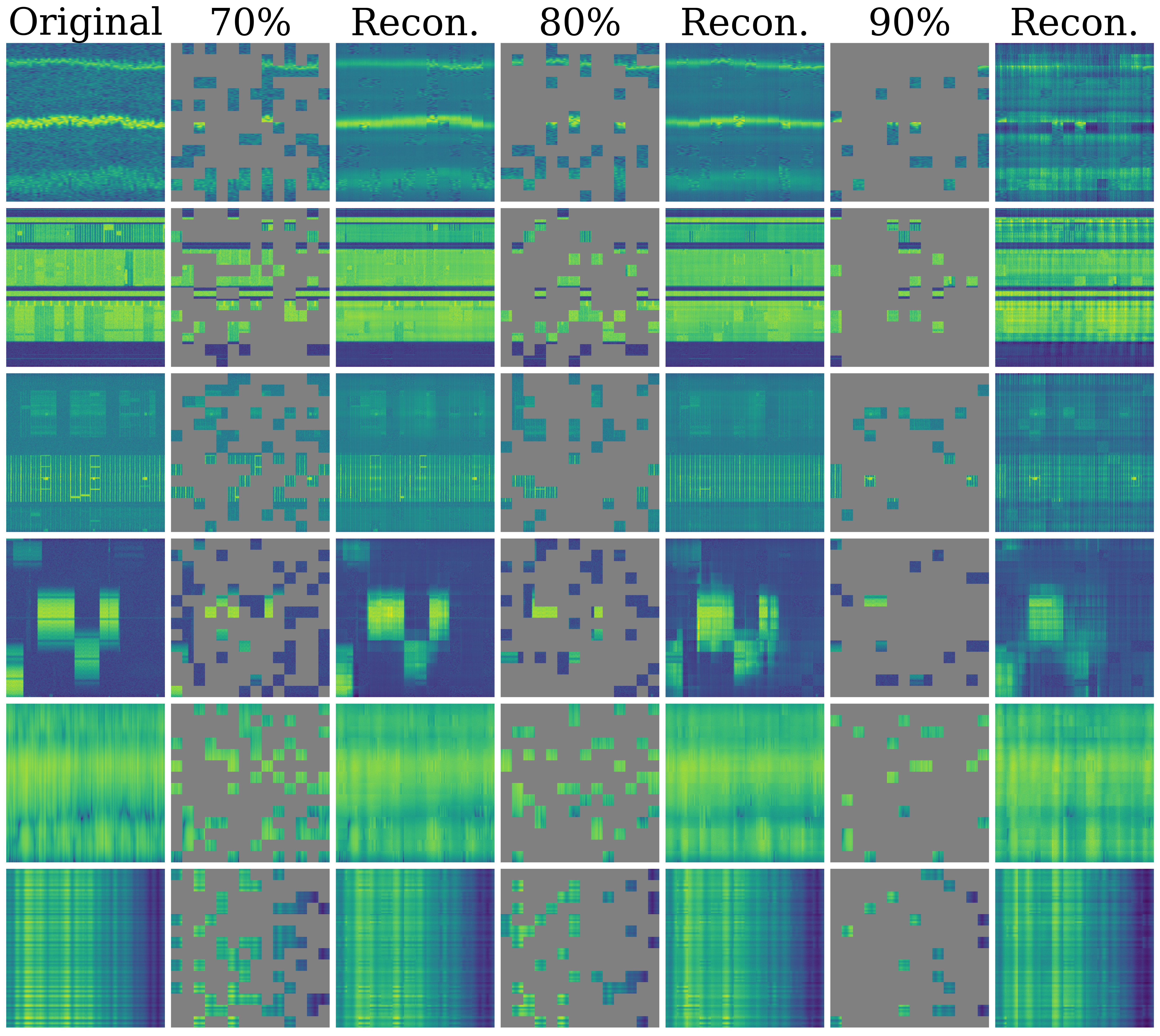}
    \caption{Reconstruction Examples for ViT-All at $70\%$, $80\%$ and $90\%$ masking ratios.}
    \label{fig:recon_examples}
\end{figure}

We begin by examining the performance of ViT‑All, pre‑trained on all the three pre-training datasets from Section \ref{sec:datasets}. Figure \ref{fig:recon_examples} shows reconstruction examples from ViT-All, illustrating that its reconstructed images are closely aligned with the originals. Due to its extensive pre-training, we expect ViT-All to deliver the strongest overall performance. We apply conventional fine‑tuning on each downstream task and evaluate its performance.

Next, we analyze the impact of pre-training data on downstream task performance from two perspectives: 1) the size and diversity of the pre-training dataset, and 2) the correlation between pre-training data and downstream tasks. 

In addition to ViT-All, we pre-train four other ViT models on distinct datasets. The first model, ViT-RFS, is pre-trained on the RF-S dataset; the second, ViT-WiFi, is pre-trained on the WiFi-CSI dataset; the third, ViT-5G, is pre-trained on the 5G-CSI dataset; 
\begin{table}[h!]
\centering
\caption{Model Hyper-parameters}
\label{tab:vit_small}
\begin{tabular}{lcc}
\toprule
\textbf{Hyperparameter}       & \textbf{ViT Encoder} & \textbf{ViT Decoder} \\
\midrule
Patch Size                    & \multicolumn{2}{c}{$16$}             \\
Blocks                        & $12$               & $8$                \\
Embed Dim.                    & $512$              & $256$              \\
Hidden Dim.                   & $2048$             & $1024$             \\
Attention Heads               & $8$                & $16$               \\
Parameters (M)                & $38$               & $7$                \\
\bottomrule
\end{tabular}
\end{table}
\begin{table}[h!]
\centering
\caption{Pre-training and Fine-tuning Hyper-parameters}
\label{tab:training_hp}
\begin{tabular}{lcc}
\toprule
\textbf{Parameter} & \textbf{Pre-training} & \textbf{Fine-tuning} \\
\midrule
Batch Size         & \multicolumn{2}{c}{$256$}                      \\
Learning Rate (LR) & \multicolumn{2}{c}{$0.001$}                    \\
Optimizer          & \multicolumn{2}{c}{Adam}                     \\
LR Scheduler       & \multicolumn{2}{c}{Cosine Annealing with linear warm-up} \\
Weight Decay       & \multicolumn{2}{c}{$0.05$}                     \\
Layer Decay        & --  & $0.75$                 \\
Epochs             & $800$ & $200$                  \\
Warm-up Epochs      & $40$  & $10$                   \\
Mask Ratio         & $0.75$ & --                  \\     
\bottomrule
\end{tabular}
\end{table}
\begin{figure*}[h!]
    \centering
    \begin{subfigure}{0.45\linewidth}
        \centering
        \includegraphics[width=\linewidth, keepaspectratio]{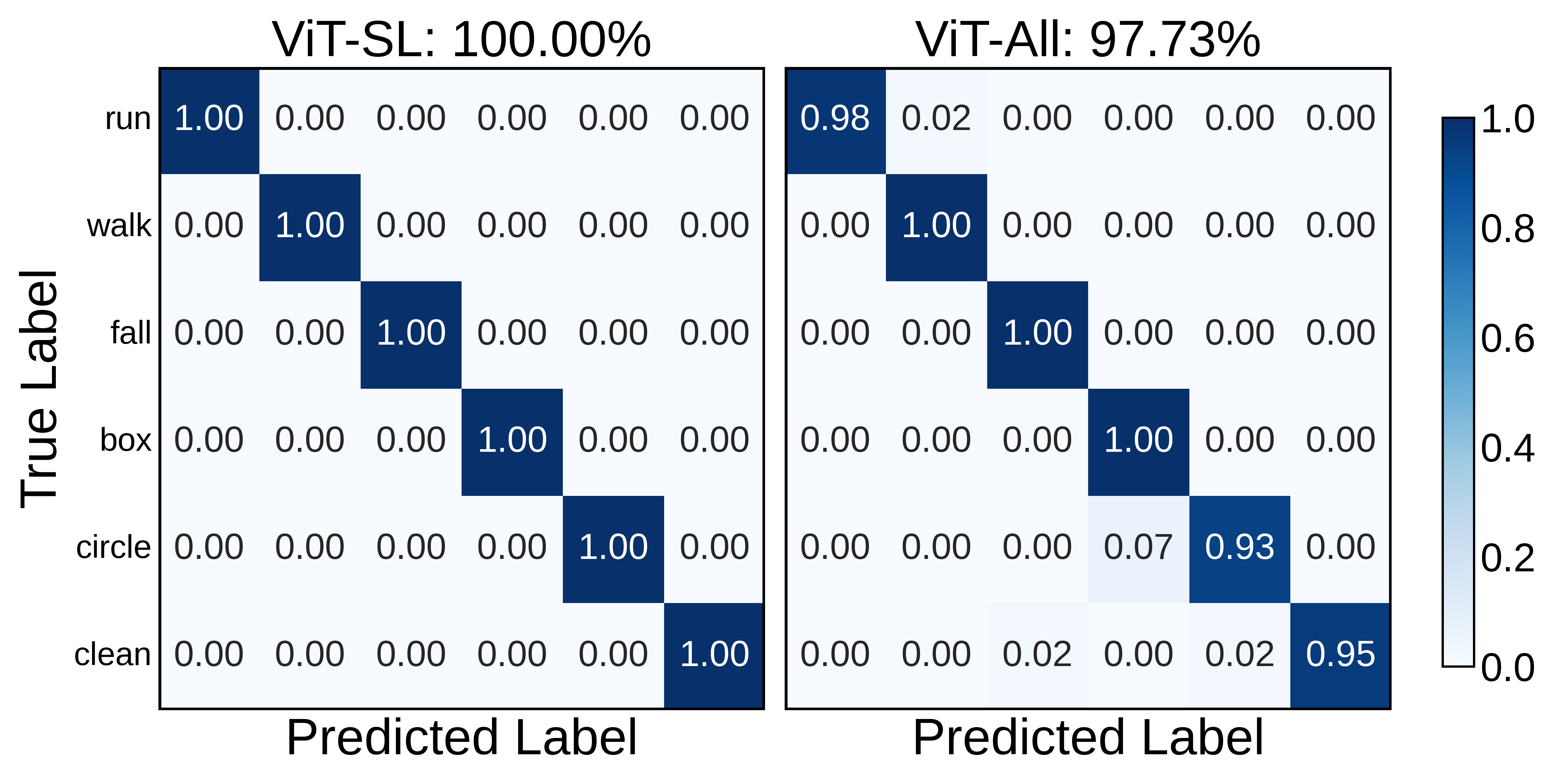}
        \captionsetup{justification=centering}
        \caption{Human Activity Sensing}
        \label{fig:best_human_sensing}
    \end{subfigure}
    \begin{subfigure}{0.45\linewidth}
        \centering
        \includegraphics[width=\linewidth, keepaspectratio]{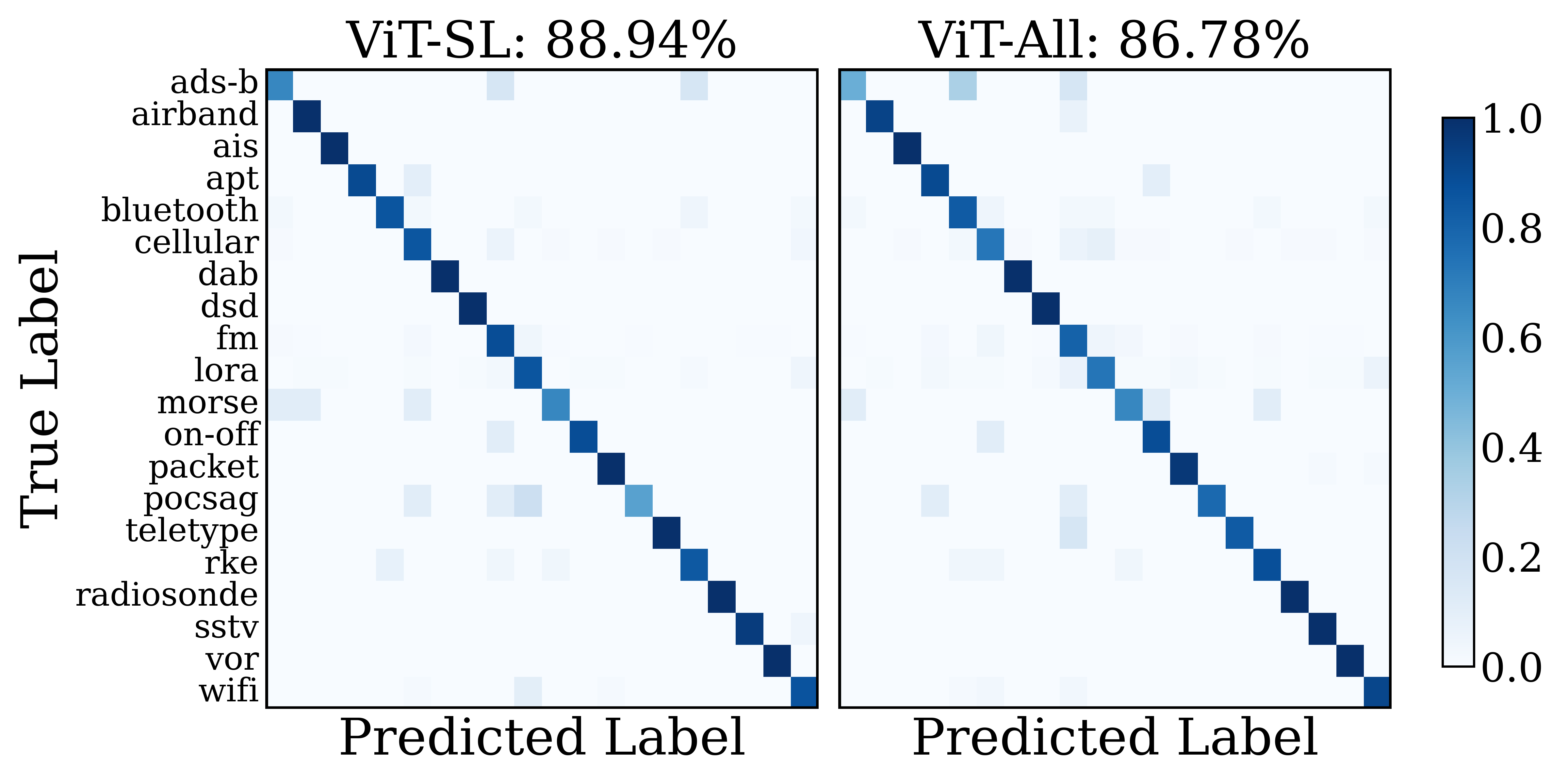}
        \captionsetup{justification=centering}
        \caption{RF Signal Classification}
        \label{fig:best_rfsig}
    \end{subfigure}
    \begin{subfigure}{0.45\linewidth}
        \centering
        \includegraphics[width=0.8\linewidth, keepaspectratio]{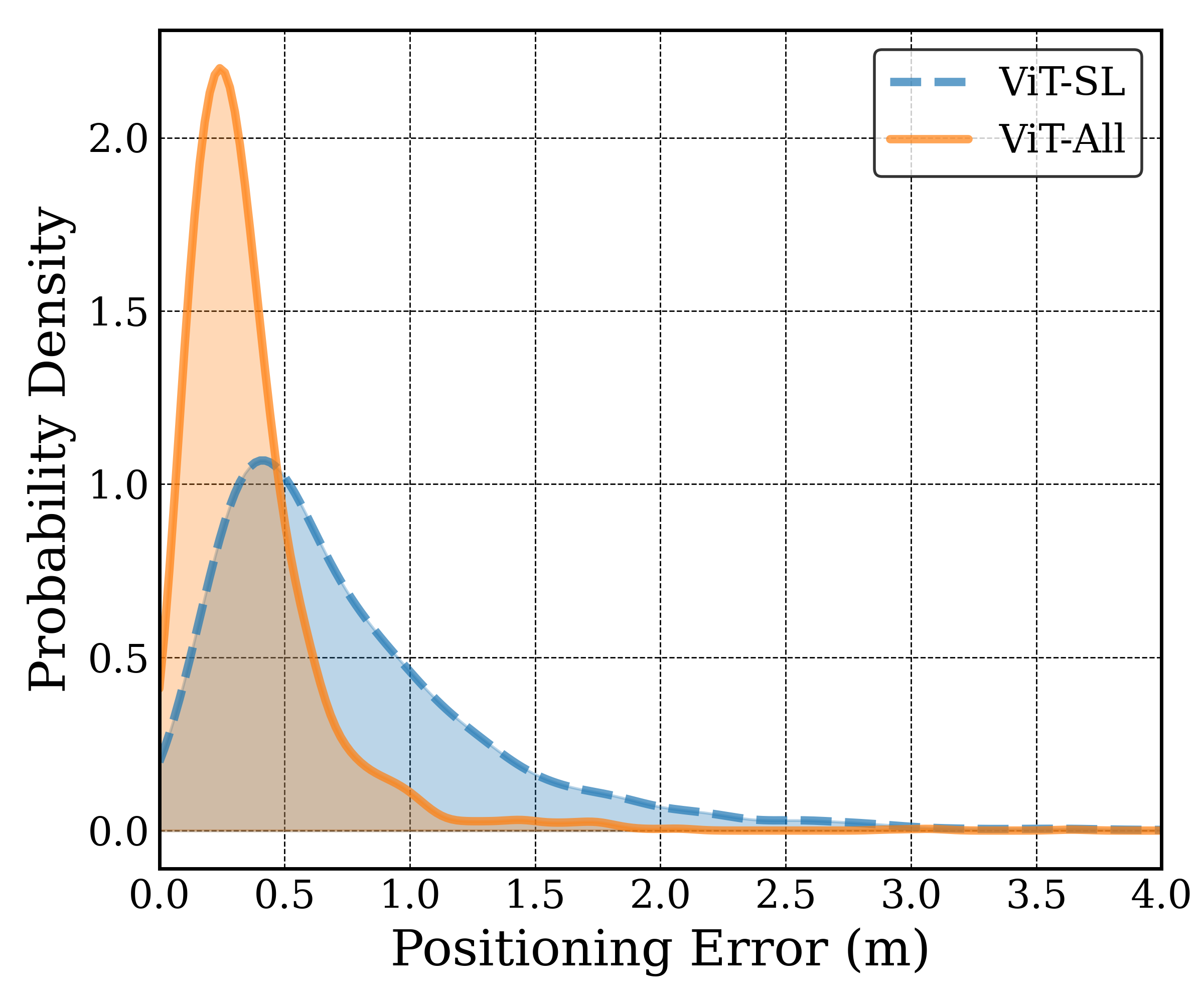}
        \captionsetup{justification=centering}
        \caption{5G NR Positioning}
        \label{fig:best_positioning}
    \end{subfigure}
    \begin{subfigure}{0.45\linewidth}
        \centering
        \includegraphics[width=0.88\linewidth, keepaspectratio]{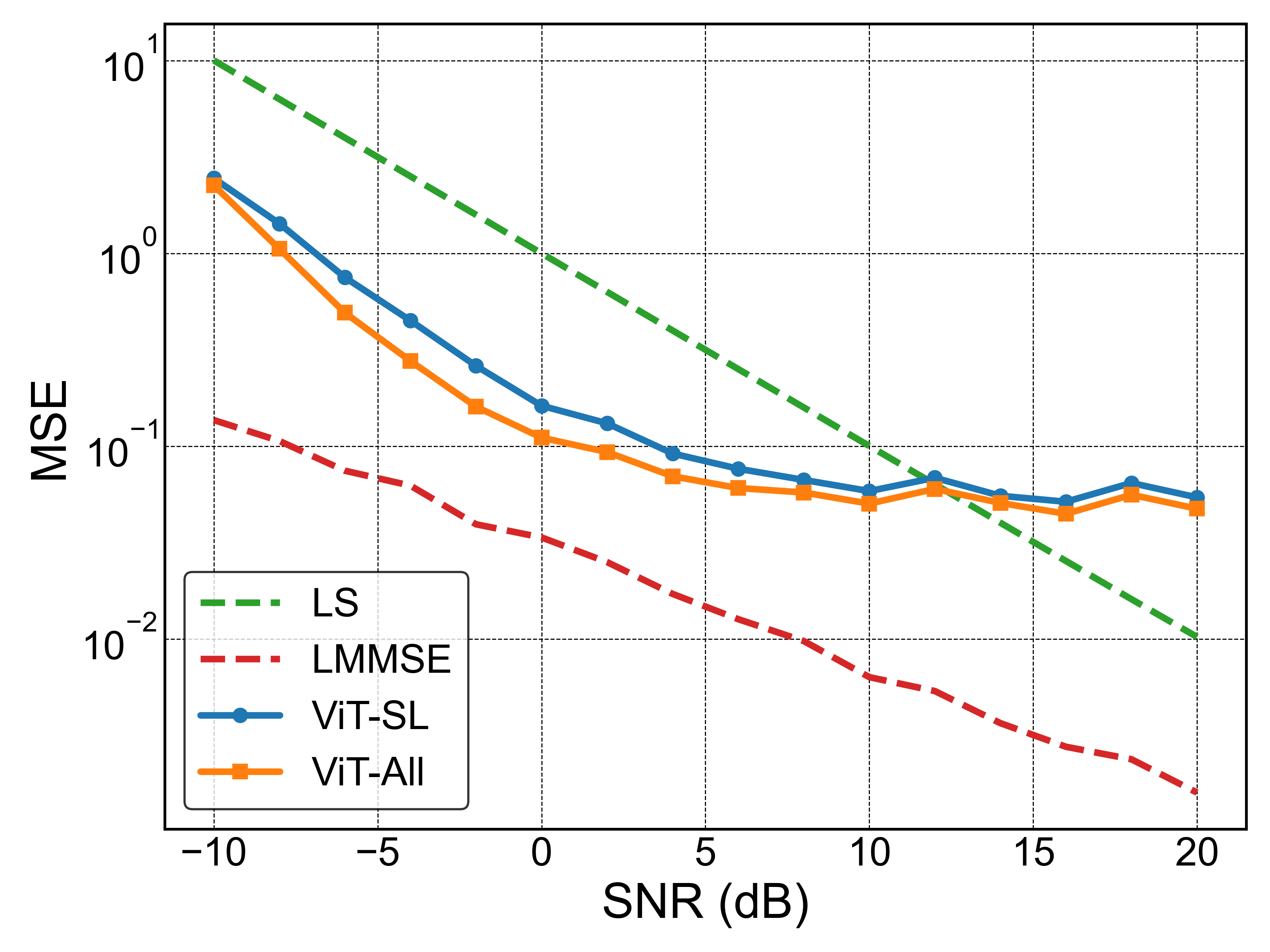}
        \captionsetup{justification=centering}
        \caption{MIMO-OFDM Channel Estimation}
        \label{fig:best_ce}
    \end{subfigure}
    \caption{ViT-All Performance on Downstream Tasks}
    \label{fig:vit_all_performance}
\end{figure*}
and the fourth, ViT-RFS/WiFi, is pre-trained on the combination of the RF-S and WiFi-CSI dataset. The configuration of ViT models is shown in Table \ref{tab:vit_small}. 

We then fine-tune each pre-trained model on all downstream tasks and evaluate their performance. 
This setup is designed to assess the individual contributions of each pre-training dataset and determine which tasks benefit most from them, thereby quantifying the utility of the correlation between pre-training data and downstream tasks. 
The supervised baseline, ViT-SL, enables us to assess the benefits of our pre-training strategy relative to learning directly on the downstream tasks. Table \ref{tab:training_hp} details the pre-training and fine-tuning hyper-parameters.

Finally, we assess the performance improvements achieved with LoRA fine-tuning compared to conventional fine-tuning. In this comparison, we highlight that LoRA requires considerably fewer task-specific parameters to reach a similar or even better level of performance. 

\subsection{ViT-All Performance Evaluation}

We compare ViT‑All’s fine‑tuning performance against ViT‑SL across the four downstream tasks. For human activity sensing, the backbone remains fully frozen and only the task head is fine‑tuned; for the other three tasks, we unfreeze the last $2$ of $12$ ViT blocks, retaining $80\%$ parameter sharing, and fine‑tune them alongside the task head. Performance is evaluated via confusion matrices for human activity sensing and RF signal classification, the probability density of positioning error for 5G NR positioning, and mean squared error across SNR for channel estimation. Figure \ref{fig:vit_all_performance} summarizes these results, with confusion matrices in Figures \ref{fig:best_human_sensing} and \ref{fig:best_rfsig}, the positioning error density in Figure \ref{fig:best_positioning}, and channel estimation error versus SNR in Figure \ref{fig:best_ce}

Overall, the WFM demonstrates strong generalization to downstream tasks, outperforming ViT-SL in both 5G NR positioning and MIMO-OFDM channel estimation while maintaining competitive results for human activity sensing and RF signal classification. In particular, the improvement for 5G NR positioning is substantial: the average distance error is reduced by half, coupled with a lower standard deviation, as illustrated in Figure \ref{fig:best_positioning}.

\begin{table*}[h!]
\centering
\begin{threeparttable}
\caption{Performance Metrics for Each Pre-training Model on Downstream Tasks}
\renewcommand{\arraystretch}{1.05}

\label{tab:pretraining_study}
\begin{tabular}{l|cc|cc|ccc|c|}
\hline
Model & \multicolumn{2}{c|}{Human Activity Sensing} & \multicolumn{2}{c|}{RF Signal Classification} & \multicolumn{3}{c|}{5G NR Positioning} & \multicolumn{1}{c|}{Channel Estimation} \\
\cline{2-9}
& Accuracy\tnote{*} (\%) & Epochs\tnote{\dag} & Accuracy\tnote{*} (\%) & Epochs\tnote{\dag} & Mean Error (m) & Std. Dev. (m) & Epochs\tnote{\dag} & Mean Square Error \\
\hline
ViT-RFS      & $89.29 \pm 5.00$ & $122$ & $86.13 \pm 1.16$ & $141$ & $1.19 \pm 0.06$ & $0.85$ & $186$ & $0.413 \pm 0.031$\\
ViT-WiFi     & $96.16 \pm 0.60$ & $20$  & $85.54 \pm 1.72$ & $114$ & $0.49 \pm 0.03$ & $0.38$ & $179$ & $0.422 \pm 0.021$\\
ViT-5G       & $79.33 \pm 5.86$ & $123$ & $66.83 \pm 2.72$ & $163$ & $0.83 \pm 0.01$ & $0.65$ & $190$ & $0.374 \pm 0.033$\\
ViT-All      & $95.67 \pm 1.93$ & $82$  & $86.05 \pm 0.79$ & $128$ & $0.41 \pm 0.02$ & $0.33$ & $182$ & $0.329 \pm 0.015$\\
ViT-RFS/WiFi & $94.86 \pm 2.23$ & $118$ & $85.24 \pm 1.62$ & $117$ & $0.40 \pm 0.02$ & $0.36$ & $176$ & $0.363 \pm 0.028$\\
ViT-SL       & $98.32 \pm 0.92$ & $107$ & $88.07 \pm 2.30$ & $91$  & $0.81 \pm 0.05$ & $0.63$ & $182$ & $0.405 \pm 0.009$\\
\hline
\end{tabular}
\begin{tablenotes}
\scriptsize
\item[*] \textit{Accuracy} is computed as the mean per-class accuracy, which provides a more balanced evaluation in the presence of class imbalance.
\item[\dag] \textit{Epochs} represent the number of fine-tuning epochs required to reach convergence.
\end{tablenotes}
\end{threeparttable}
\end{table*}

\begin{figure*}[h!]
\centering
\begin{subfigure}{0.4\linewidth}
    \centering
    \includegraphics[width=\linewidth, keepaspectratio]{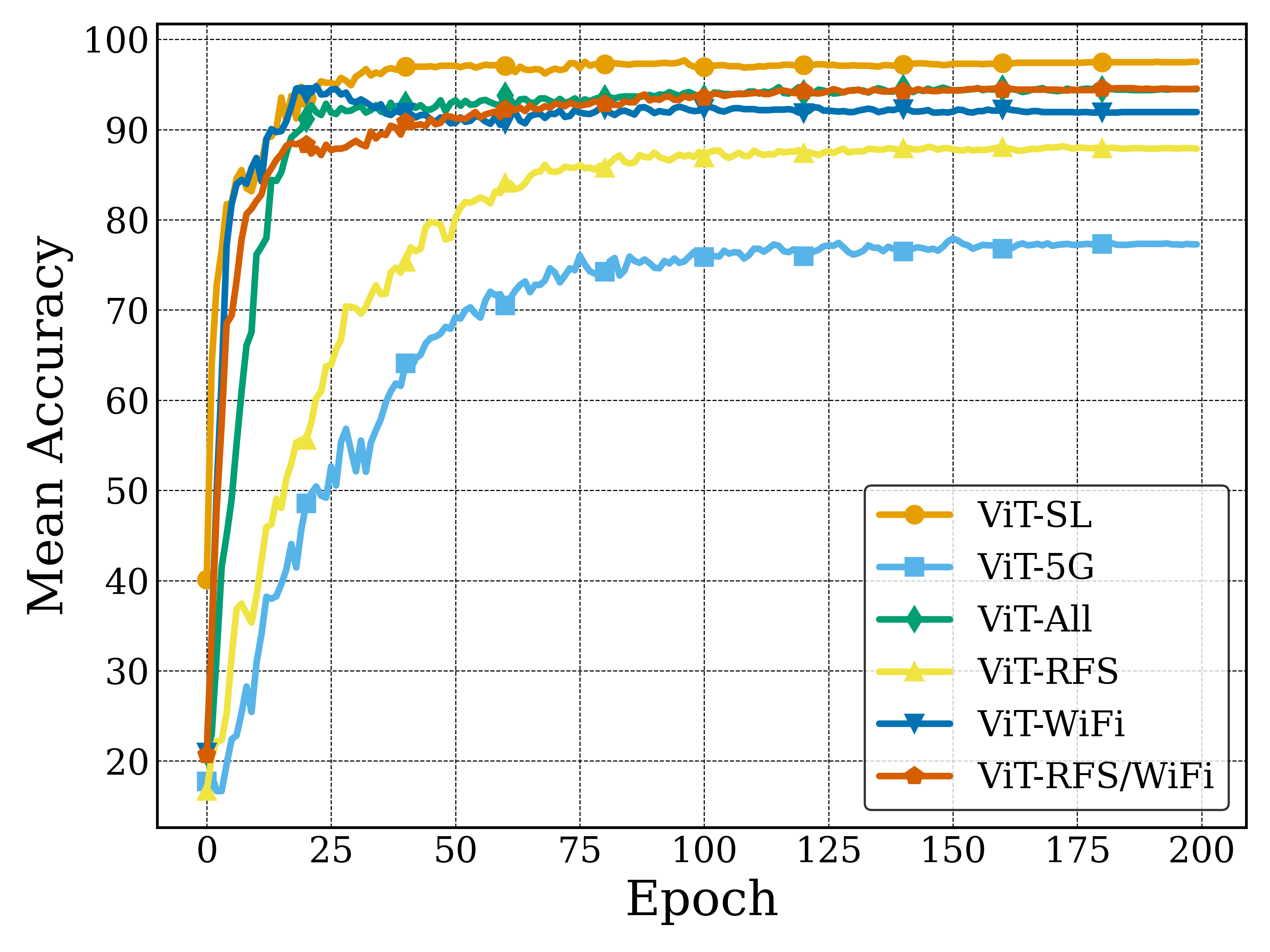}
    \captionsetup{justification=centering}
    \caption{Human Activity Sensing}
    \label{fig:human_sensing_performance_vs_epochs}
\end{subfigure}
\hskip 1cm
\begin{subfigure}{0.4\linewidth}
    \centering
    \includegraphics[width=\linewidth, keepaspectratio]{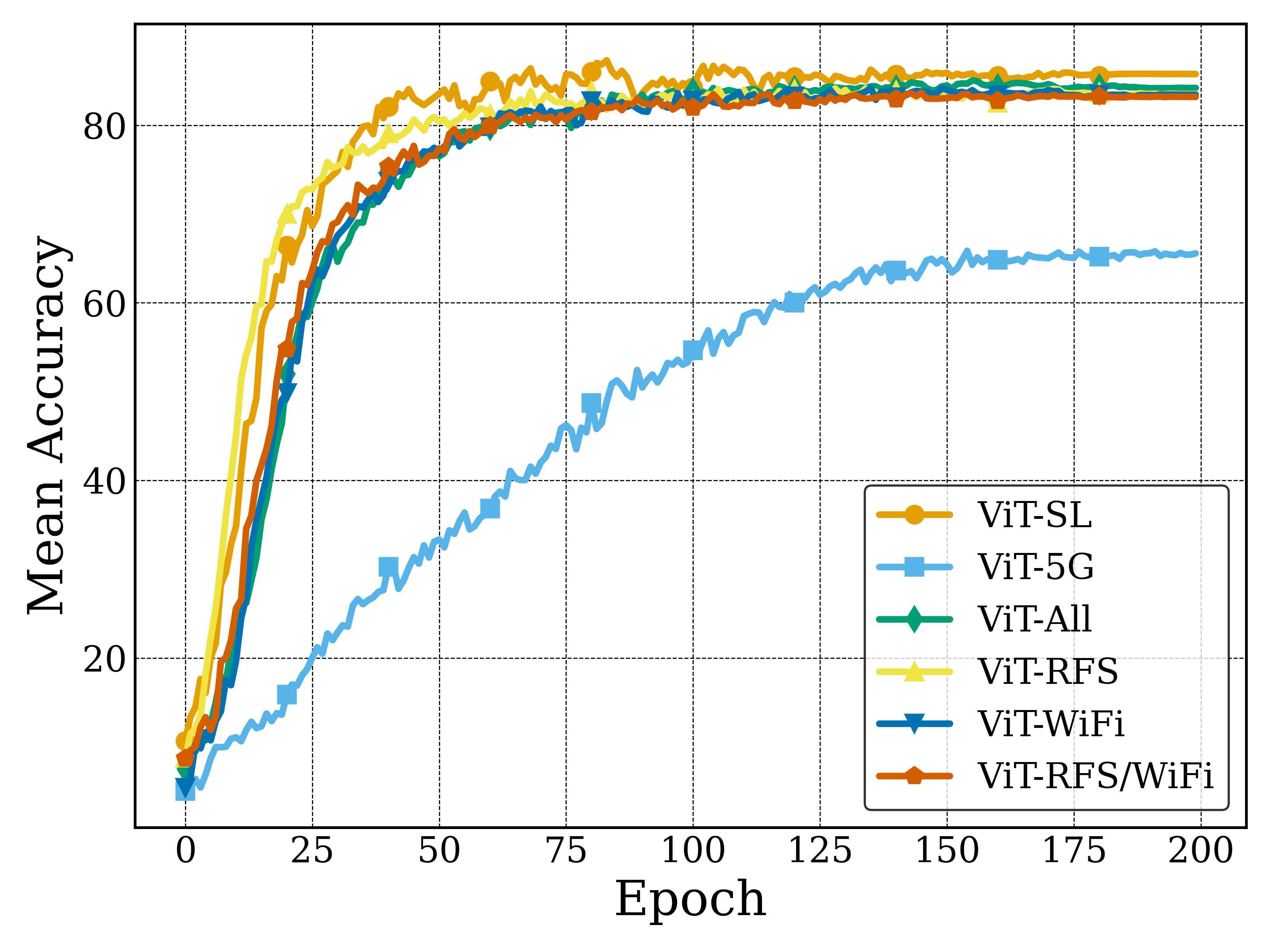}
    \captionsetup{justification=centering}
    \caption{RF Signal Classification}
    \label{fig:rfsig_performance_vs_epochs}
\end{subfigure}
\begin{subfigure}{0.4\linewidth}
    \centering
    \includegraphics[width=\linewidth, keepaspectratio]{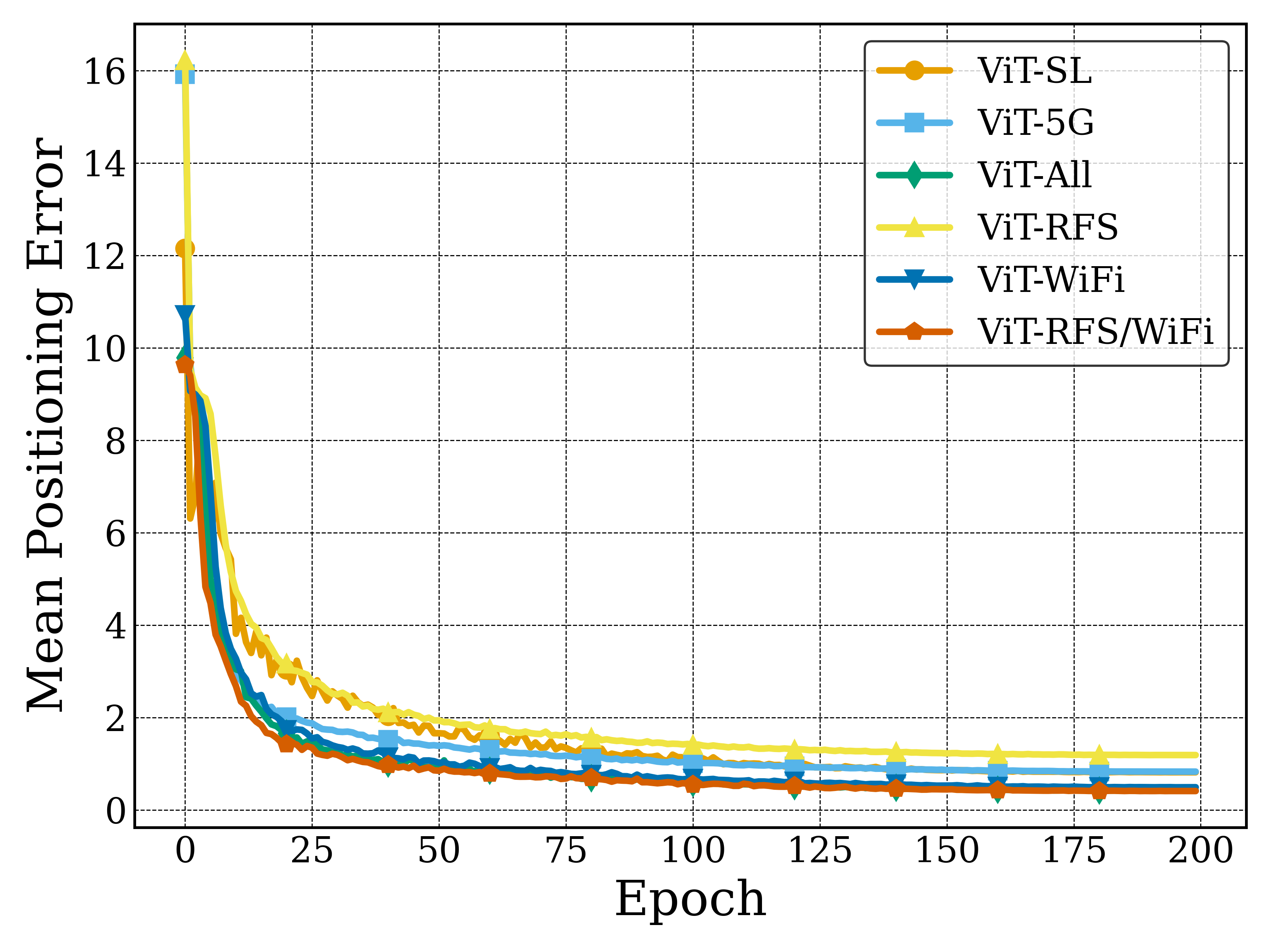}
    \captionsetup{justification=centering}
    \caption{5G NR Positioning}
    \label{fig:positioning_performance_vs_epochs}
\end{subfigure}
\hskip 1cm
\begin{subfigure}{0.4\linewidth}
    \centering
    \includegraphics[width=\linewidth, keepaspectratio]{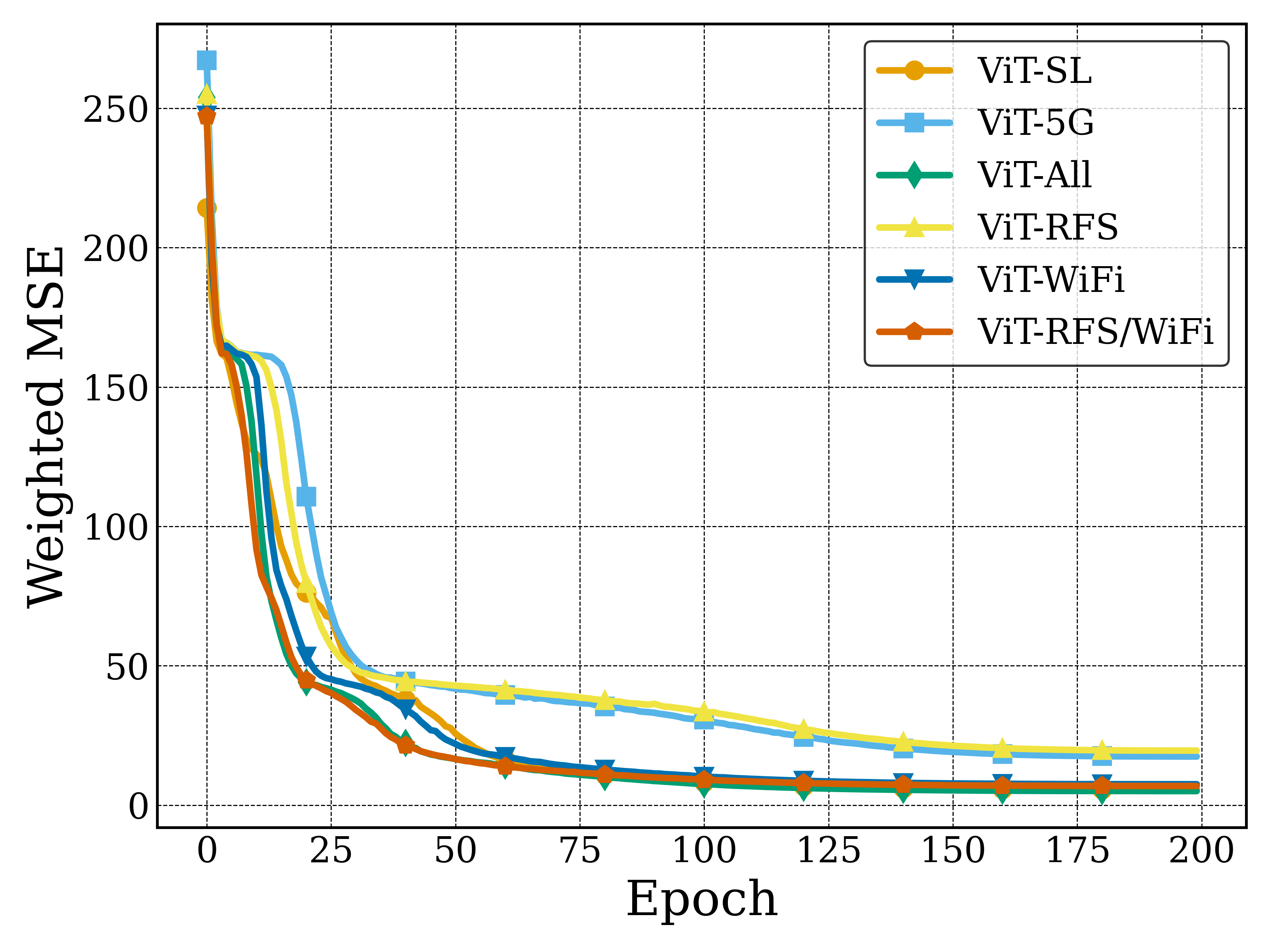}
    \captionsetup{justification=centering}
    \caption{MIMO-OFDM Channel Estimation}
    \label{fig:channel_estimation_performance_vs_epochs}
\end{subfigure}
\caption{Performance trends over epochs for all downstream tasks.}
\label{fig:performance_vs_epochs}
\end{figure*}

For MIMO-OFDM channel estimation, besides ViT-SL, we also utilize the least-squares (LS) and Linear Minimum Mean Square Error (LMMSE) estimators as additional baselines. Like the DL models, LS uses transmitted and received pilots to estimate the channel. LMMSE requires additional knowledge of the channel’s statistical properties through time and frequency covariance matrices which is likely impractical in real-world deployments. Hence, LMMSE serves as an idealized best-case baseline.
On average, both ViT-All and ViT-SL outperform LS, with significant gains in the SNR range from $-10$ to $5$ dB. However, their performance saturates as the SNR increases, with LS surpassing them for SNRs above $13$ dB. In practice, we can implement a back-off strategy, allowing the system to revert to the simpler LS estimator at high SNR values.

We attribute the saturation to the nature of optimization, which tends to prioritize higher errors and; hence, lower SNRs. By employing the SNR-weighted MSE loss defined in equation \eqref{eq:loss_snr_mse}, we successfully delayed the onset of saturation, but it did not fully eliminate the effect. Further research is needed to develop training strategies that effectively handle a wide range of SNRs. A similar phenomenon was reported in \cite{ce_iotais}, where under-sampling low-SNR signals was used to prevent them from dominating the optimization.


\subsection{Impact of Pre-training Data}

In this experiment, we evaluate the impact of pre-training data on model performance. To ensure a fair evaluation, we hold all fine-tuning settings for each task constant across different models, except for optimization parameters, such as the learning rate, which we vary to realize the best performance. Each model is fine-tuned with an equal number of frozen blocks and an identical task head configuration. As in the previous experiment, for human activity sensing the pre‑trained backbone remains fully frozen, whereas for the other three tasks, we unfreeze the last $2$ of $12$ blocks and fine‑tune them alongside the task head.

As evaluation metrics, we use mean per‑class accuracy for human activity sensing and RF signal classification (to address class imbalance); mean positioning error for 5G NR positioning (the average Euclidean distance between predicted and true positions); and mean squared error for channel estimation (comparing channel estimates with the target). Additionally, we track the number of epochs required to reach peak performance as a measure of convergence speed. Table \ref{tab:pretraining_study} presents the ablation study results averaged over $7$ runs. For brevity, the number of epochs required for the channel estimation models to converge is omitted, as they all converge towards the end of training. Furthermore, Figure \ref{fig:performance_vs_epochs} illustrate the performance on the test set for every downstream task, as a function of epochs during fine-tuning.

Our evaluation reveals that fine-tuning performance is strongly tied to the alignment between pre-training data and downstream task characteristics. For instance, ViT-WiFi, which leverages highly correlated pre-training data to human activity sensing (both being WiFi-CSI of very close nature), demonstrates rapid convergence by reaching peak performance in an average of $20$ epochs. Similarly, ViT-RFS, pre-trained on data quite aligned with RF signal classification (both being spectrograms, albeit from different sources), exhibits the steepest performance ascent among all pre-trained models, as shown in Figure \ref{fig:rfsig_performance_vs_epochs}.

In contrast, ViT-5G generally under-performs on downstream tasks, with the exception of MIMO-OFDM channel estimation, where it ranks a close third. This discrepancy is likely due to a mismatch between the distribution of the pre-training data and the downstream tasks. The strong performance in MIMO-OFDM channel estimation is to be expected, given that the 5G CSI data closely resembles the MIMO-OFDM transmissions used in this task, as confirmed by visual inspection of Figures \ref{fig:pretrain_5g_csi} and \ref{fig:mimo-ofdm}. What is surprising, however, is its poor performance in 5G NR positioning, despite this task also operating on 5G CSI. A closer examination suggests that the 5G-CSI dataset differs in nature from the 5G NR positioning dataset, likely due to variations in hardware, environmental conditions, and data pre-processing methods, which can be observed in Figures \ref{fig:pretrain_5g_csi} and \ref{fig:positioning}.

Regarding data size, we observe that larger pre-training datasets clearly improve performance in 5G NR positioning and MIMO-OFDM channel estimation, as demonstrated by the ViT-All and ViT-RFS/WiFi models, while the benefits are less clear for human activity sensing and RF signal classification. This indicates that although larger datasets can be advantageous, using domain-irrelevant pre-training data may occasionally lead to negative knowledge transfer.

\subsection{LoRA Fine-tuning}

Given the performance gap between ViT-SL and conventional fine-tuning on RF signal classification, which requires fine‑tuning two full ViT blocks to achieve reasonable accuracy, we explore LoRA fine-tuning as a more parameter-efficient alternative. Our goal is to evaluate the trade-off between the number of trainable parameters and performance, while keeping the WFM frozen.

\begin{figure}[t!]
    \centering
    \includegraphics[width=0.9\linewidth, keepaspectratio]{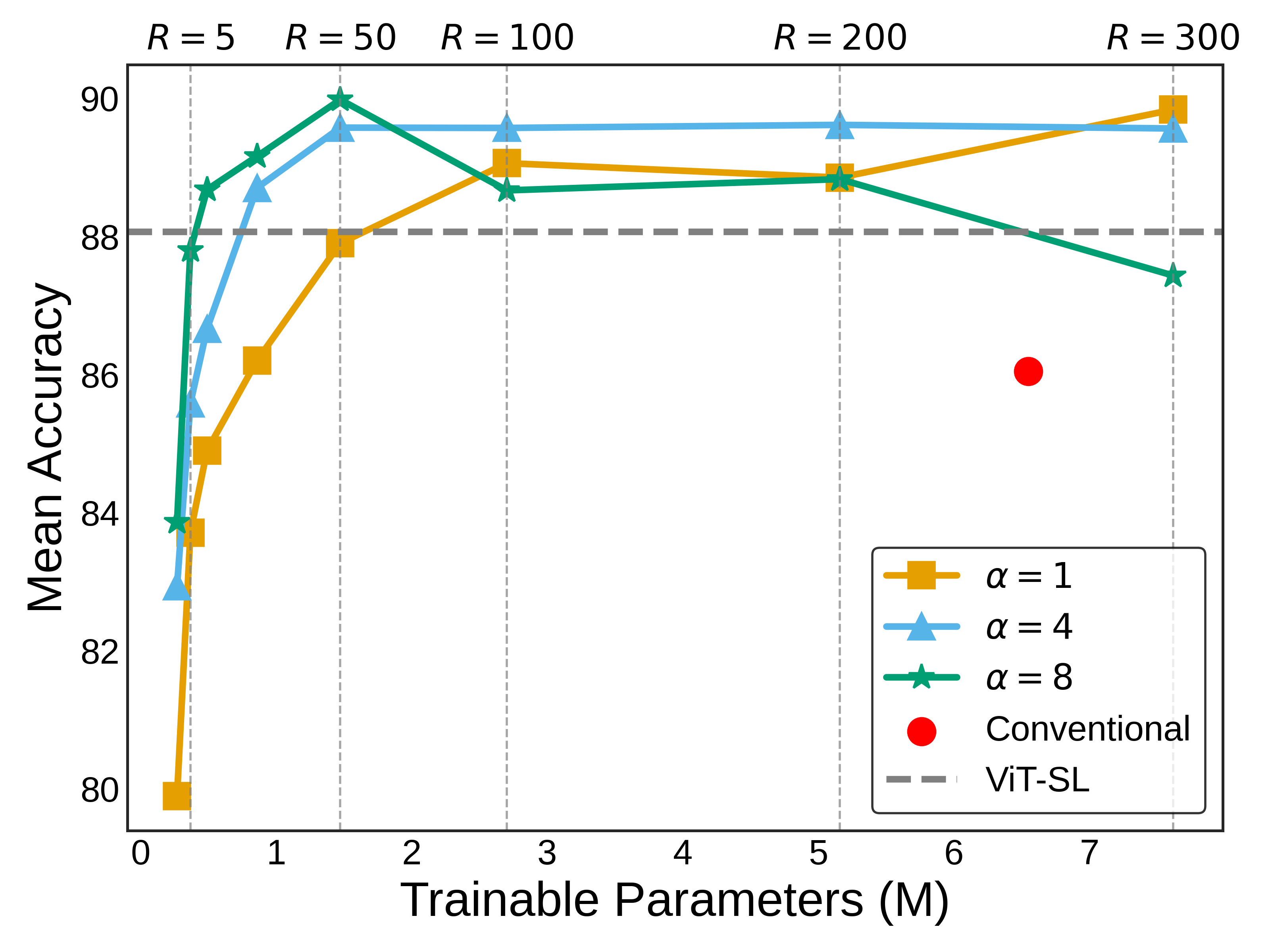}
    \caption{LoRA Pareto Curve for RF Signal Classification}
    \label{fig:lora_pareto}
\end{figure}
\begin{figure}[h!]
    \centering
    \includegraphics[width=\linewidth, keepaspectratio]{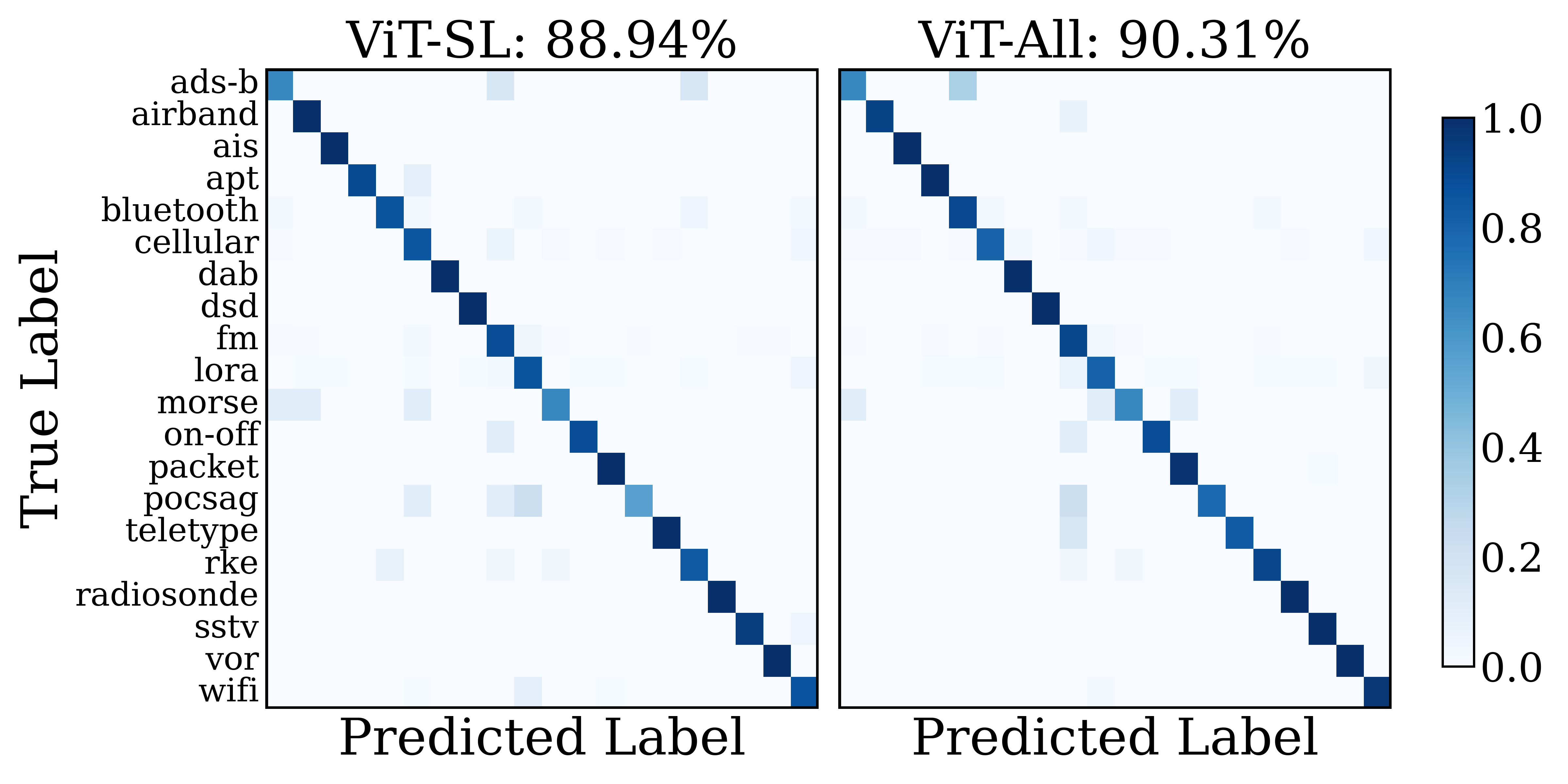}
    \captionsetup{justification=centering}
    \caption{Performance of LoRA fine-tined ViT-ALL compared to ViT-SL on RF signal classification.}
    \label{fig:best_rfsig_lora}
\end{figure}

We fine-tune ViT-All using various settings for the LoRA rank $R$, which determines the number of trainable parameters, and scaling factor $\alpha$, which controls the strength of the LoRA adaptation. 

The results in Figure \ref{fig:lora_pareto}, averaged across three runs, indicate that for large values of $\alpha$, LoRA models can outperform both conventional fine‑tuning and ViT‑SL while adding less than $1$ million additional parameters. 
\noindent{The best performance is achieved at $\alpha = 8$ and $R = 50$ which adds approximately $1.5$ million additional parameters.}
On the other hand, increasing the rank at high values of $\alpha$ can degrade performance, suggesting that the model may begin to unlearn useful knowledge. In contrast, for smaller values like $\alpha = 1$, performance improves steadily with rank, though the gains are more modest compared to higher $\alpha$ settings. Figure \ref{fig:best_rfsig_lora} shows the confusion matrices for the LoRA fine‑tuned ViT‑All ($\alpha=8, R=50$) versus ViT‑SL, highlighting ViT‑All’s superior accuracy.

\section{Conclusion}
\label{sec:conclusion}

In this work, we introduced WavesFM, the first Wireless Foundation Model (WFM), capable of directly processing image-like wireless modalities, such as spectrograms and CSI, and IQ signals to support a diverse range of sensing, communication and localization tasks. Our proposed architecture combines a shared ViT-based backbone with task-specific MLP heads, and incorporates LoRA for parameter-efficient fine-tuning. This design promotes full parameter sharing across tasks, significantly reducing the computational and memory footprint without sacrificing performance.

Through extensive experiments on four downstream tasks: human activity sensing, RF signal classification, 5G NR positioning, and MIMO-OFDM channel estimation, we demonstrated the strong generalization capabilities of our WFM. Compared to supervised baselines trained individually, our approach achieves superior performance while sharing $80\%$ of the parameters across tasks. We also show that using domain-relevant pretraining data not only improves performance but also accelerates convergence, reducing training time by up to 5× in some cases. The results demonstrate that a unified WFM can support a variety of tasks and also offer significant improvements in performance and efficiency. Our approach provides a new direction for shaping the architecture and functionality of upcoming 6G networks. 

\balance
\bibliography{bibliography.bib}
\bibliographystyle{ieeetr}

\end{document}